\documentclass[twocolumn]{aastex701}
\usepackage{multirow}
\usepackage{graphicx}
\usepackage{array}
\usepackage{bbding}
\usepackage[shortlabels]{enumitem}
\usepackage{xspace}

\newcommand{\ohcue}{\ensuremath{\mathrm{(O/H)_{Cue}}}\xspace}
\newcommand{\ohstrong}{\ensuremath{{\mathrm{(O/H)_{Cue, strong}}}}\xspace}
\newcommand{\ohte}{\ensuremath{\mathrm{(O/H)}_{T_e}}\xspace}
\newcommand{\ohfull}{\ensuremath{\mathrm{(O/H)_{fid}}}\xspace}
\newcommand{\ohsubset}{\ensuremath{\mathrm{(O/H)_{subset}}}\xspace}

\newcommand{\yp}{LBT $Y_P$ Project\xspace}
\newcommand{\ang}{\ensuremath{\mathrm{\AA}}\xspace}

\usepackage{enumitem}
\usepackage{hyperref}
\usepackage{CJK}

\begin{document}
\begin{CJK*}{UTF8}{gbsn}
\title{CECILIA: Using Flexible Photoionization Models to Accurately Measure Abundances without Auroral Lines}

\correspondingauthor{Nathalie Korhonen Cuestas}
\author[orcid=0000-0003-2385-9240,gname=Nathalie,sname=Korhonen Cuestas]{Nathalie A. Korhonen Cuestas}
\affiliation{Department of Physics and Astronomy, Northwestern University, 2145 Sheridan Road, Evanston, IL, 60208, USA}
\affiliation{Center for Interdisciplinary Exploration and Research in Astrophysics (CIERA), Northwestern University, 1800 Sherman Avenue, Evanston, IL, 60201, USA}
\email[show]{nathaliekorhonencuestas2029@u.northwestern.edu} 

\author[orcid=0000-0001-6369-1636,gname=Allison,sname=Strom]{Allison L. Strom}
\affiliation{Department of Physics and Astronomy, Northwestern University, 2145 Sheridan Road, Evanston, IL, 60208, USA}
\affiliation{Center for Interdisciplinary Exploration and Research in Astrophysics (CIERA), Northwestern University, 1800 Sherman Avenue, Evanston, IL, 60201, USA}
\email{allison.strom@northwestern.edu} 

\author[orcid=0000-0002-0682-3310,gname=Yijia,sname=Li]{Yijia Li (李轶佳)}
\affiliation{Center for Interdisciplinary Exploration and Research in Astrophysics (CIERA), Northwestern University, 1800 Sherman Avenue, Evanston, IL, 60201, USA}
\email{yijia.li@northwestern.edu} 

\author[orcid=0000-0002-0361-8223,gname=Noah,sname=Rogers]{Noah S. J. Rogers}
\affiliation{Center for Interdisciplinary Exploration and Research in Astrophysics (CIERA), Northwestern University, 1800 Sherman Avenue, Evanston, IL, 60201, USA}
\email{noah.rogers@northwestern.edu} 

\author[orcid=0000-0002-6034-082X,gname=Caroline,sname=von Raesfeld]{Caroline L. von Raesfeld}
\affiliation{Department of Physics and Astronomy, Northwestern University, 2145 Sheridan Road, Evanston, IL, 60208, USA}
\affiliation{Center for Interdisciplinary Exploration and Research in Astrophysics (CIERA), Northwestern University, 1800 Sherman Avenue, Evanston, IL, 60201, USA}
\email{CarolinevonRaesfeld2027@u.northwestern.edu} 

\author[orcid=0000-0002-8459-5413,gname=Gwen,sname=Rudie]{Gwen C. Rudie}
\affiliation{The Observatories of the Carnegie Institution for Sciences, 813 Santa Barbara Street, Pasadena, CA 91101, USA}
\email{gwen@carnegiescience.edu} 

\author[orcid=0000-0002-6967-7322,gname=Ryan,sname=Trainor]{Ryan F. Trainor}
\affiliation{Department of Physics and Astronomy, Franklin \& Marshall College, 637 College Avenue, Lancaster, PA 17603, USA}
\affiliation{William H. Miller III Department of Physics and Astronomy, Johns Hopkins University, Baltimore, MD 21218, USA}
\email{ryan.trainor@fandm.edu} 

\author[orcid=0009-0008-2226-5241,gname=Menelaos,sname=Raptis]{Menelaos Raptis}
\affiliation{Department of Astrophysical Sciences, Princeton University, 4 Ivy Lane, Princeton, NJ 08544, USA}
\affiliation{Department of Physics and Astronomy, Franklin \& Marshall College, 637 College Avenue, Lancaster, PA 17603, USA}
\email{mr0194@princeton.edu} 

\author[orcid=0000-0002-1945-2299,gname=Zhuyun,sname=Zhuang]{Zhuyun Zhuang}
\affiliation{Center for Interdisciplinary Exploration and Research in Astrophysics (CIERA), Northwestern University, 1800 Sherman Avenue, Evanston, IL, 60201, USA}
\email{zhuyun.zhuang@northwestern.edu} 


\begin{abstract}

JWST has facilitated the detection of faint, auroral emission lines needed to directly measure the electron temperature and elemental abundances in high-redshift galaxies. However, such measurements still require substantial observational investment, potentially making self-consistent, population-level studies of direct-method metallicity scaling relations across multiple epochs of Cosmic time prohibitively expensive. Photoionization modeling could allow us to measure metallicity using more routinely-observed emission lines across larger galaxy samples with shallower spectra. We use $z\sim2$ galaxies from the CECILIA survey and local galaxies and H~II regions from the \yp to compare abundance measurements made via the $T_e$ method to abundances estimated using the novel photoionization modeling code, \texttt{Cue}. Based on the results, we explore the viability of, biases introduced by, and the best practices for measuring gas-phase metallicity with photoionization modeling. We show that \texttt{Cue} can be used to accurately measure O/H while remaining agnostic to the ionizing spectrum and that accurate recovery of O/H can be achieved with relatively bright oxygen and sulfur lines, demonstrating that \texttt{Cue} could be used to measure metallicity in large samples that lack the depth necessary for direct measurements of O/H. However, using sulfur emission lines can also severely bias the results of photoionization modeling in high-redshift galaxies, where non-Solar abundance patterns are common. We make recommendations regarding the use of photoionization models to measure metallicity in local and high-redshift samples and future developments to photoionization modeling methods. 


\end{abstract}



\section{Introduction}
\par The chemical enrichment of a galaxy reflects the interplay of inflows, outflows, and star formation over time. Population-level trends linking the metallicity of the interstellar medium (ISM), stellar mass ($M_\star$), star formation rate (SFR), relative elemental abundances, and other nebular properties therefore reflect the evolution and growth of galaxies. The shape, scatter, and evolution of abundance scaling relations such as the mass-metallicity relation \citep[MZR, e.g.,][]{2004Tremonti, 2006Lee, 2012Berg} and fundamental metallicity relation \citep[FMR, e.g.,][]{2010Mannucci, 2013Andrews, 2020Curti} are sensitive to both changes in the strength and enrichment of inflows and outflows \citep{2004Tremonti, 2008Finlator, 2012Dave, 2024Bassini} and typical star formation histories \citep[SFH,][]{1979Tinsley, 2024Garcia, 2025Marszewski}. Accurate estimation of the intrinsic scatter can help us to understand systematic differences in the evolution of galaxies across cosmic time, \citep[e.g.,][]{2022Strom, 2025Korhonen, 2026Kotiwale} but this requires robust, unbiased abundance measurements across large galaxy samples. 
\par Bulk gas-phase metallicity is typically traced by the abundance of oxygen relative to hydrogen (O/H or, simply, metallicity in this letter). Gas-phase metallicity can be measured via four methods: the recombination line (RL) method, the ``direct" (or $T_e$) method, strong-line calibrations, and photoionization modeling. 
\par The RL method provides a very direct probe of elemental abundances as the strength of metal RLs relative to a hydrogen RL is highly insensitive to variations in nebular temperature and density (which collisionally excited lines depend upon) and directly sensitive to the relevant ionic abundance. However, such RLs are extremely faint, making them challenging to detect in nearby objects and inaccessible in observations of distant galaxies. 
\par The $T_e$ method uses temperature- and density-sensitive line ratios to determine the emissivity of a given transition and then estimate an ionic abundance. The method requires the detection of faint, temperature-sensitive auroral lines and the ability to resolve closely-spaced density-sensitive line doublets. While the $T_e$ method is often seen as the ``gold standard" for measuring O/H (and other abundances) in distant galaxies, it is not free of assumptions, each of which may introduce some systematic uncertainty. For example, unless all relevant ionization zones are observed, the temperature of unseen zones is typically estimated using empirical or theoretical $T_e$-$T_e$ relations \citep[e.g.,][]{1990Stasinska, 2003PerezMontero, 2004Garnett, 2021Rogers}, which have substantial scatter and vary in form. Moreover, studies of local H~II regions and planetary nebulae have long observed a discrepancy of $\sim0.2-0.3$ dex between metallicity estimates from the $T_e$ and RL methods \citep{1983French, 2005GarciaRojas, 2007GarciaRojas}. The exact origin of the abundance discrepancy factor (ADF) is debated, but it has been suggested that the $T_e$ method is biased towards the lowest-metallicity, highest-temperature (and therefore most luminous) regions of the nebula, resulting in a metallicity estimate that is lower than the average metallicity of the nebula \citep{1967Peimbert, 2023MendezDelgado}.
\par Prior to the launch of JWST, detection of faint auroral lines in high-redshift galaxies was limited to small samples of lensed galaxies \citep[e.g.,][]{2009Yuan, 2011Rigby, 2012Christensen, 2014James}, individual highly star-forming galaxies \citep[SFGs, e.g.,][]{2016Sandersb, 2023Sanders_a}, and stacked spectra \citep[e.g.,][]{2016Steidel, 2023Clarke}. To circumvent the need for faint auroral line detections, strong-line calibrations \citep[e.g.,][among many others]{1994McGaugh, 2004Pettini} have historically been used to estimate O/H in larger samples of high-$z$ galaxies with more typical spectral depth \citep[e.g.,][]{2006Erb, 2014Steidel, 2018Sanders, 2021Topping, 2025Korhonen}. 
\par Strong-line ratios are readily observed but are also sensitive to other physical parameters such as ionization, density, the relative abundance of other elements, and the ionizing spectrum. As a result, strong-line calibrations are inaccurate when applied to samples where the physical conditions are dissimilar to those in the calibration sample. Recent work using high-$z$ galaxies as the calibration sample \citep[][]{2024Laseter, 2025Cataldi, 2025Chakraborty, 2025Sanders, 2025Scholte} can go some way towards mitigating the problems inherent to applying local calibrations to high-redshift samples. However, they still suppress significant scatter in the inferred O/H due to intrinsic scatter in a line ratio at fixed O/H driven by variations in the ionizing spectrum \citep{2014Steidel, 2018Strom, 2025Shapley}, ISM ionization \citep{2002Kewley, 2014Steidel}, and N/O \citep{2014Perez-Montero, 2016Sanders}. 
\par Photoionization models generally combine a library of ionizing spectra, for example from stellar population synthesis or active galactic nucleus (AGN) models, and a radiative transfer code to predict the strength of different emission lines given a set of gas parameters. Then, the optimal solution can be found either using Bayesian methods or $\chi^2$-minimization. Various frameworks exist to make photoionization model-based abundance measurements. They vary in terms of their treatment of multiple-element abundances, choice of ionizing spectrum, treatment of dust, and fitting algorithms. For example, some codes treat N/O as a free parameter (e.g., \texttt{BOND}, \citealt{2016ValeAsari}; \texttt{GalDNA}, \citealt{2018Strom}; \texttt{HOMERUN}, \citealt{2024Marconi}; and \texttt{Cue}, \citealt{2025Li}) while others tie it explicitly to the oxygen abundance \citep[e.g., \texttt{H~II-CHI-MISTRY}, \citealt{2014Perez-Montero}; \texttt{IZI}][]{2015Blanc}. Abundances from photoionization models tend to be higher than $T_e$ measurements, especially at lower metallicities, with typical offsets ranging from 0.17 to 0.56 dex, depending on the modeling framework \citep{2014Perez-Montero, 2015Blanc, 2016ValeAsari}, potentially due to the same temperature inhomogeneities that drive the ADF between $T_e$ and RL methods. Unlike strong-line calibrations, photoionization modeling frameworks that consider a range of physical conditions do not suppress the intrinsic scatter of a strong-line ratio at fixed metallicity, which is driven by variations in ionization, density, relative abundances, and the ionizing spectrum.
\par Photoionization modeling is a promising avenue for making self-consistent measurements of metallicity in large, statistical samples as it neither relies on similar physical conditions in a calibration sample nor does it necessarily require a suite of faint auroral emission lines. However, a careful examination of the performance of novel photoionization modeling methods across samples with different elemental abundance patterns is still required. In this letter, we use galaxies with detections of multiple auroral and nebular emission lines and detailed multi-element abundances from the CECILIA program ($z\sim2$ SFGs) and the \yp (local metal-poor galaxies and H~II regions) to directly compare the results of the $T_e$ method and photoionization modeling, using \texttt{Cue} \citep{2025Li}. Based on the results, we outline best practices for using \texttt{Cue} and other photoionization models to measure metallicity in samples with different abundance patterns, spectroscopic depth, and wavelength coverage. 
\par The letter is organized as follows: the CECILIA and \yp samples are introduced in Sections \ref{sec:cecilia} and \ref{sec:yp}, respectively. We briefly compare the two in Section \ref{sec:comparison}. Our use of \texttt{Cue} is described in Section \ref{sec:cue}, and we compare the results of \texttt{Cue} to $T_e$ metallicities in Sections \ref{sec:nonsolar} and \ref{sec:yp_cue}, with specific attention paid to the effect of multi-element abundance patterns. In Section \ref{sec:typicalspec}, we explore how \texttt{Cue} performs on more typical spectral data and which emission lines are most important in constraining the metallicity. Finally, we end by making recommendations regarding the application of \texttt{Cue} to other galaxy samples and future developments in photoionization modeling (Section \ref{sec:recs}). 

\par Throughout this paper, we assume a $\Lambda$CDM cosmology with $H_0=70\, \mathrm{km\,s^{-1}\, Mpc^{-1}}, \, \Omega_\Lambda=0.7$, and $\Omega_{m}=0.3$ and adopt the same abundance pattern assumed by \texttt{Cue}: a total oxygen abundance of $12+\log(\mathrm{O/H})_{\odot, \mathrm{total}} = 8.93$ \citep{1989Anders}, assuming a dust depletion factor of $\log(D_\mathrm{O})=-0.22$, gas-phase $\log(\mathrm{S/O})_{\odot}=-1.50$ and $\log(\mathrm{Ar/O})_\odot=-2.15$ \citep{2000Dopita}, although we note that all $T_e$-based results are insensitive to the adopted Solar scale. Throughout the text, we refer to the emission lines using their wavelengths in air, in units of $\mathrm{\AA}$ngstroms. 

\section{Samples and Data}\label{sec:data}
\subsection{CECILIA}\label{sec:cecilia}
\par CECILIA is a Cycle 1 JWST program \citep[PID 2593,][]{2021Strom} targeting UV-color-selected SFGs at $1\lesssim z\lesssim3$ designed to get detailed information on the chemical abundances of galaxies at Cosmic Noon. Galaxies in the CECILIA survey were selected from the Q2343 field of the Keck Baryonic Structure Survey \citep[KBSS;][]{2010Steidel, 2012Rudie, 2017Strom}. The CECILIA survey adds ultra-deep (29.5 hr) G235M/F170LP and 1.1 hr G395M/F290LP JWST/NIRSpec observations to existing Keck/MOSFIRE spectra and extensive multi-band photometry. The survey design, NIRSpec spectral reduction, emission line measurements, dust correction, and spectral energy density (SED)-fitting have been discussed extensively by \cite{2023Strom, 2024Rogers, 2026Rogers_a} and we refer the reader to these works for further detail. A discussion of the spectral reduction, emission-line measurements, and SED-fitting for the larger KBSS sample can be found in \cite{2014Steidel, 2017Strom, 2025Korhonen}. The derivation of electron density ($n_e$), electron temperature ($T_e$), and O/H can be found in \cite{2026Rogers_a}.
\par The CECILIA sample consists of 33 galaxies. In this letter, we exclude galaxies with poorly-constrained SED continua (4 Ly$\alpha$-selected galaxies), galaxies at $z<2$ (4 galaxies), and Q2343-BX391, as MSA shutter failures resulted in 80\% of exposures being lost for that object. We additionally require H$\alpha$ to have been detected in NIRSpec (i.e., it did not fall in the chip gap), excluding a further 3 galaxies, and we exclude Q2343-RK120 due to poorly constrained cross-band slit loss corrections \citep[see discussion in][]{2026Rogers_a}. After removing these galaxies, we are left with a sample of 21 galaxies. 
\subsubsection{Calculating Line Luminosities}\label{sec:normlines}
\par Emission lines for CECILIA galaxies are measured across Keck/MOSFIRE $J, \,H$ and $K$, and JWST/NIRSpec G235M/F170LP and G395M/F290LP. Calculating line luminosities on a common scale therefore requires cross-band and cross-instrument normalization. Additionally, as is noted in \cite{2024Rogers}, reddening-corrected H~I recombination lines in Q2343-D40 deviate from Case B recombination, consistent with either a wavelength-dependent flux calibration error across G235M/G395M or a different attenuation curve shape. Motivated by this, we adopt a pseudo-dust reddening approach similar to that recommended by \cite{2025Stasinska} and adopted by \cite{2026Rogers_a}.
\par All lines in G235M within $5755\,\mathrm{\AA}\leq\lambda_\mathrm{rest}\leq7715\,\mathrm{\AA}$ are normalized relative to H$\alpha$. For {[N~II]}$\lambda\lambda$6548,85, He~I 6678, and {[S~II]}$\lambda\lambda$6716,31, we use line fluxes, uncorrected for the effects of dust attenuation, and for all other lines, we use dust-corrected line fluxes. We adopt the \cite{2020Reddy} dust attenuation curve and take $E(B-V)$ from \cite{2026Rogers_a}, which is calculated from the Balmer decrement, assuming H$\alpha/$H$\beta=2.82$ under Case B conditions with $T_e=1.25\times10^4$ K and $n_e=300 \, \mathrm{cm^{-3}}$ \citep{1995Storey}. When {[S~III]}$\lambda\lambda 9069,9531$ falls in G235M, we divide the uncorrected line flux by the line flux of Pa8 and then multiply this ratio by the theoretical Pa8/H$\alpha$ ratio. If {[O~III]}$\lambda\lambda$4959,5007 and H$\beta$ are available in G235M, we take the ratio of the uncorrected line fluxes and multiply it by the theoretical H$\beta$/H$\alpha$ ratio. The theoretical H~I line/H$\alpha$ ratios are calculated using the \texttt{getEmissivity} function in \texttt{pyneb} \citep[v1.1.28][]{2015Luridiana}. When available, we use the high-ionization zone $T_e$ value from \cite{2026Rogers_a} and the average of $n_e$({[S~II]}) and $n_e$({[O~II]}), or simply $n_e$({[S~II]}) if $n_e$({[O~II]}) is not available. If direct measurements are not available, we adopt fiducial values of $T_e=1.25\times10^4$ K and $n_e=300 \mathrm{\,cm^{-3}}$, reflective of typical conditions in galaxies at this redshift \citep{2016Sanders, 2018Strom}. Our calculated line luminosities depend weakly on the choice of $n_e$ and $T_e$. 
\par We normalize emission lines in G395M ({[S~III]}$\lambda\lambda9069,9531$ and He~I$\lambda$10830) relative to the closest available Paschen line. Due to the lower resolution at the blue end of the disperser, we do not use Pa8 due to possible blending. {[S~III]}$\lambda\lambda9069,9531$ is normalized relative Pa9 (which allows us to use uncorrected line fluxes) or Pa7 (using dust-corrected line intensities) when Pa9 is unavailable. He~I$\lambda$10830 is normalized relative to Pa$\gamma$ using uncorrected line fluxes. Each ratio is then multiplied by the theoretical ratio of the relevant Paschen line to H$\alpha$. 
\par For most CECILIA galaxies in our sample (14/21), {[O~III]}$\lambda\lambda$4959,5007 is only available in MOSFIRE and is correspondingly normalized using H$\beta$ in MOSFIRE. If H$\beta$ is not available in MOSFIRE, then we use dereddened and slit-loss corrected intensities and normalize relative to H$\alpha$ in MOSFIRE. We follow a similar procedure for {[O~II]}$\lambda\lambda3726,29$, which is only available in MOSFIRE, except we always use dereddened and slit-loss corrected line intensities because there is no significantly detected Balmer line close in wavelength to {[O~II]}. Additionally, three objects (Q2343-C31, Q2343-BX341, and Q2343-BX350) lack high SNR H$\alpha$ and H$\beta$ detections in MOSFIRE (although both are detected in NIRSpec, allowing us to calculate the Balmer decrement). In this case, we take the ratio of dereddened and slit-loss corrected {[O~II]}$\lambda\lambda3726,29$ and {[O~III]}$\lambda5007$ fluxes in MOSFIRE and use the {[O~III]}$\lambda5007$/H$\alpha$ ratio from NIRSpec to normalize {[O~II]} to H$\alpha$. 
\par To convert all line fluxes from a scale relative to H$\alpha$ to absolute luminosities, we multiply by the dust-corrected H$\alpha$ line intensity and 4$\pi D_L^2$ where $D_L$ is the luminosity distance, determined from the spectroscopic redshift. 

\subsection{\yp}\label{sec:yp}
\par In addition to CECILIA galaxies, we test \texttt{Cue}'s performance on local galaxies and H II regions from the Large Binocular Telescope (LBT) Primordial He Abundance Project \citep[hereafter the \yp,][]{2026Skillman}. The sample contains both metal-poor and extreme emission line objects, with spectral coverage across the optical and NIR. We specifically make use of emission lines detected using the LBT Multi-Object Double Spectrograph \citep[MODS,][]{2010Pogge}. Details regarding the spectral reduction and emission line measurements can be found in \cite{2026Rogers_b}. Thanks to the wavelength coverage (3300-10000 \ang), resolution ($R\sim1850$ in the blue channel and $R\sim2300$ in the red channel), and depth of these spectra, the electron temperature, density, and elemental abundances can be calculated for multiple ions, giving us a detailed picture of the physical conditions in \yp objects \citep{2026Rogers_b}. The \yp sample contains 62 local galaxies and H II regions. We exclude objects where shocks (SHOC133 and WJ1205+4551) and density-bounded nebulae (Mrk 71) are suspected, as \texttt{Cue}'s framework does not accurately describe these systems. 
\par To calculate the luminosity of each emission line, we adopt a similar pseudo-dust reddening approach to what is described in Section \ref{sec:normlines}. We do not have to account for cross-instrument calibration in this case, so we simply select the nearest well-detected H~I line as the reference line. As was done for CECILIA galaxies, we then normalize the line relative to H$\alpha$ using the theoretical H~I line/H$\alpha$ ratio. Instead of the \cite{2020Reddy} attenuation curve, we adopt the \cite{1989Cardelli} attenuation curve, which is better suited to local galaxies and consistent with the treatment of \yp objects in other works \citep{2026Skillman, 2026Rogers_b}. We exclude lines affected by blending ({[Ne~III]$\lambda3967$} and He~I$\lambda3888$) or telluric absorption (primarily {[S~III]$\lambda9531$}). Lastly, we convert from flux relative to H$\alpha$ to luminosity by multiplying by the dust-corrected H$\alpha$ line intensity and $4\pi D^2$, where $D$ is the distance measured using the tip of the red giant branch method for LeoP \citep{2013McQuinn} and Leoncino \citep{2020McQuinn}, and retrieved from CosmicFlows4 \citep{2020Kourkchi} for all other objects. 

\subsection{Comparison of CECILIA and \yp objects}\label{sec:comparison}
\par CECILIA is a sample of typical SFGs at Cosmic Noon, while \yp objects are local metal-poor, extreme emission line galaxies and nebulae. The two samples allow us to test \texttt{Cue}'s performance across a wide range of metallicities, abundance patterns, and nebular properties. 
\begin{figure}
    \centering
    \includegraphics[width=1\linewidth]{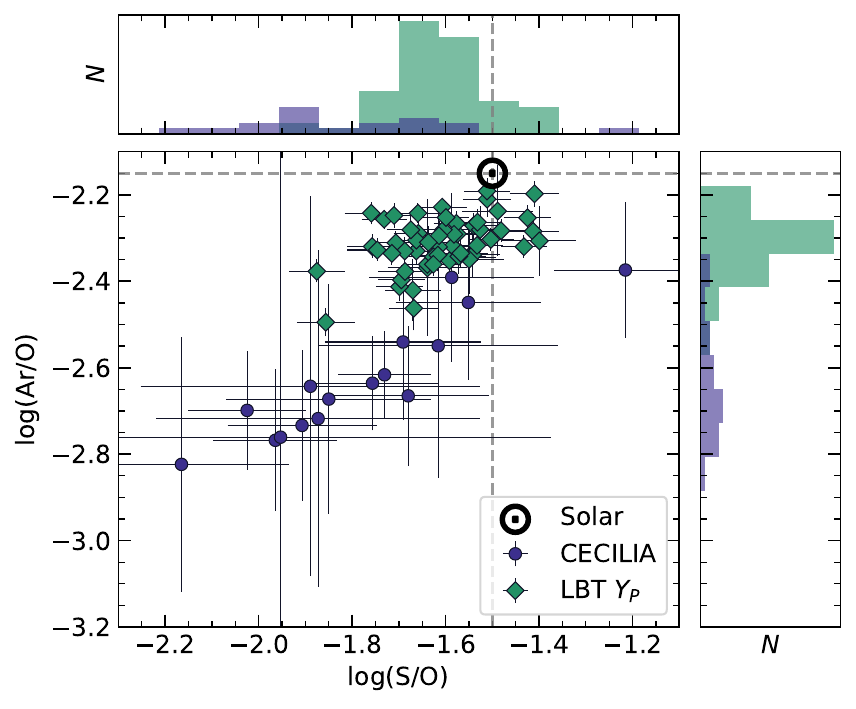}
    \caption{CECILIA galaxies (purple circles) tend to have lower S/O (x-axis) and Ar/O (y-axis) than \yp objects (green diamonds), which are more consistent with the Solar abundance values adopted in \texttt{Cue} \citep[][gray dashed lines]{2000Dopita}. Marginal distributions in S/O and Ar/O are shown in the horizontal and vertical histograms.}
    \label{fig:yp_cecilia_ar_s}
\end{figure}
\par A key difference between the two samples is their enrichment histories. CECILIA galaxies have enrichment histories dominated by core-collapse supernovae (CCSNe), resulting in subsolar Ar/O and S/O abundances \citep[see purple points in Figure \ref{fig:yp_cecilia_ar_s} and][]{2026Rogers_a}. \yp objects, on the other hand, have been enriched over longer timescales, allowing for more Type Ia enrichment, resulting in Ar/O and S/O abundances consistent with Solar (see green points in Figure \ref{fig:yp_cecilia_ar_s}). We note that while $Y_P$ galaxies are offset to slightly subsolar values relative to the \cite{2000Dopita} scale adopted by \texttt{Cue}, the offset is minimal when compared to other commonly used Solar abundance scales \citep[e.g.,][]{2021Asplund}. The combination of the two samples allows us to assess the impact of multi-element abundance patterns on the results of photoionization modeling. 
\begin{figure}
    \centering
    \includegraphics[width=1 \linewidth]{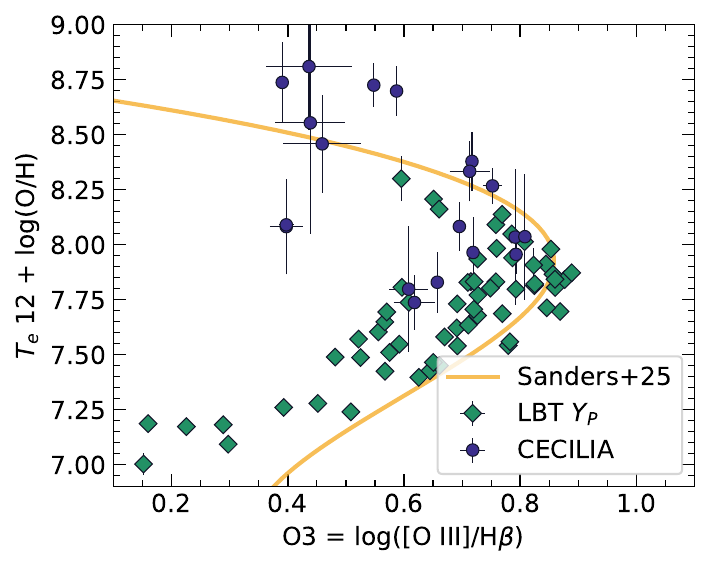}
    \caption{O3 (x-axis) is double-valued in O/H (y-axis), as is reflected by many strong-line calibrations \citep[here, the high-$z$ recalibration from][is shown in yellow to guide the eye]{2025Sanders}. CECILIA galaxies (purple circles) populate the high-metallicity branch and the turnover region while \yp objects (green diamonds) populate the lower-metallicity branch and turnover region.}
    \label{fig:yp_cecilia_r3}
\end{figure}
\par The combination of samples also allows us to populate both the high- and low-metallicity branches of strong-line ratios, such as O3 and R23.\footnote{$\mathrm{O3\equiv\log\left(\frac{[O\,III]\lambda5007}{H\beta}\right)}$, \\$\mathrm{R23\equiv\log\left(\frac{[O\,II]\lambda\lambda3726,29 + [O\,III]\lambda\lambda4959,5007}{H\beta}\right)}$} Figure \ref{fig:yp_cecilia_r3} shows how O3 varies as a function of O/H; at $12+\log(\mathrm{O/H})\lesssim7.75$, increasing O/H boosts O3 by increasing the number of available O$^{++}$ ions, while at $12+\log(\mathrm{O/H})\gtrsim8.25$, increasing O/H leads to a lower electron temperature, thereby suppressing O3. In the turnover region ($12+\log(\mathrm{O/H})\sim8.0$), O3 is largely insensitive to changes in O/H and variations in O3 may instead be driven by changes in the ionization parameter and ionizing spectrum (Korhonen Cuestas et al., in prep). A similar trend is seen in R23. Combining CECILIA and \yp objects allows us to probe all three regimes.
\par Based on strong-line diagnostics, we can infer that the ionization conditions are relatively similar in \yp objects and CECILIA galaxies. While \yp objects have higher ionization parameters than typical local SFGs, elevated ionization parameters are common in the high-redshift universe \citep{2008Brinchmann, 2009Hainline, 2014Nakajima, 2014Steidel, 2015Hayashi, 2023Sanders_b, 2026Cleri}. Therefore, we can explore the performance of photoionization models across a range of metallicities while keeping other nebular properties relatively consistent.
\begin{figure*}
    \centering
    \includegraphics[width=1 \linewidth]{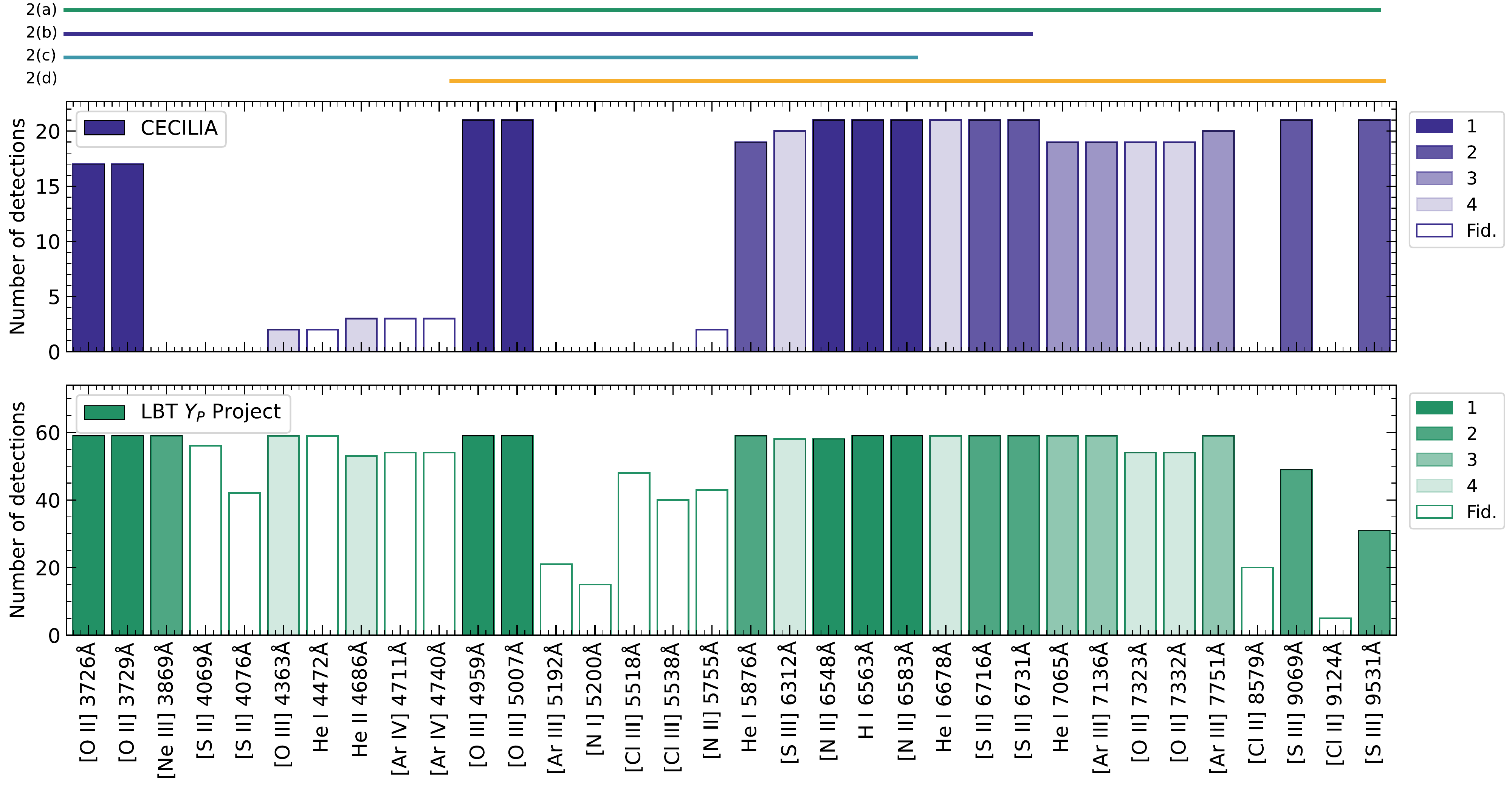}
    \caption{All of the commonly-detected lines in CECILIA galaxies are also detected in \yp objects and the emission lines span the same rest-wavelength range. \yp objects have significantly better coverage of weaker emission lines bluewards of 5800\ang. The y-axis shows the number of detections of each emission line across the CECILIA (top panel) and \yp (bottom panel) samples. The shading of each bar indicates which luminosity-based lists a given line is included in (1-4; lines only included in the fiducial run are unfilled), and the horizontal lines (2(a)-2(d)) above the panels indicate the lines included in wavelength-based lists (see Sections \ref{sec:yp_1234} and \ref{sec:yp_abcd}). Note that while {[S~III]}$\lambda\lambda9069,9531$ is detected in all \yp objects, one of the two lines is typically excluded due to telluric contamination.}
    \label{fig:line_list_summary}
\end{figure*}
\par Figure \ref{fig:line_list_summary} summarizes the emission lines used in the analysis of our two samples and shows the detection rate of each line. All emission lines used in the analysis of CECILIA galaxies are also detected in the LBT $Y_P$ sample. While {[S~III]}$\lambda\lambda9069,9531$ is detected in all \yp objects, telluric contamination makes it such that typically only one of the two lines is used in the analysis. Since the lines have a fixed flux ratio, we do not lose any information when one [S III] line is excluded. \yp objects have significantly deeper spectra than CECILIA galaxies bluewards of $\lambda_\mathrm{rest}\sim 5800\ang$, allowing us to explore the effect of fainter emission lines that are accessible for galaxies at Cosmic Noon with NIRSpec/G140M spectral coverage. 

\section{Photoionization Modeling}\label{sec:phot_model}
\subsection{\texttt{Cue}}\label{sec:cue}
\par We use \texttt{Cue} \citep{2025Li} to conduct photoionization modeling of CECILIA galaxies and \yp objects. \texttt{Cue} is distinct in its adoption of a flexible piecewise power law as the ionizing spectrum, rather than a fixed stellar population or AGN model. This approach is in contrast to typical approaches where the ionizing spectrum is modeled using a simple stellar population, such as a single burst at different ages \citep[e.g.,][]{2016ValeAsari, 2024Marconi} or a constant star formation history at different stellar metallicities \citep{2018Strom}. \texttt{Cue}'s agnosticism towards the ionizing source is well suited to modeling galaxies with extreme or composite ionizing sources \citep{2024Li, 2025Helton, 2025Wang, 2026Naidu}. Here, we assess \texttt{Cue}'s utility in measuring chemical abundances in more typical SFGs and compare its performance to the $T_e$ method.
\par \texttt{Cue} takes as input a list of dust-corrected line luminosities and upper limits and samples the posterior parameter distribution, varying both the ionizing spectrum and gas properties using dynamic nested sampling \citep[\texttt{dynesty},][]{2020Speagle, 2022Koposov}. For each proposed set of parameters, a neural net emulator of \texttt{Cloudy} \citep[v22.00,][]{2023Chatzikos} is used to calculate model line luminosities. The difference between the modeled and observed line luminosities is then used to calculate the likelihood of the proposed parameters. 

\begin{deluxetable}{c|c}
\tablecaption{Priors used in \texttt{Cue} to model the slope ($\alpha_i$) and relative strength ($F_{i+1}/F_i$) of the different ionizing spectrum sections and nebular conditions. All parameters are defined as in \cite{2025Li}. \label{tab:cuepriors}}
\tablehead{
\colhead{Ionizing Spectrum Parameter} & \colhead{Prior}
}
\startdata
$\alpha_\mathrm{He\,II}$     & $\mathcal{U}(1, 42)$  \\
$\alpha_\mathrm{O\,II}$     & $\mathcal{U}(-0.3, 30)$  \\
$\alpha_\mathrm{He\,I}$     & $\mathcal{U}(-1.1, 14)$ \\
$\alpha_\mathrm{H\,I}$     & $\mathcal{U}(-1.7, 8)$\\
$\log(F_\mathrm{O\,II}/F_\mathrm{He\,II})$ & $\mathcal{U}(-0.1, 10.1)$ \\
$\log(F_\mathrm{He\,I}/F_\mathrm{O\,II})$ & $\mathcal{U}(-0.5, 1.9)$ \\
$\log(F_\mathrm{H\,I}/F_\mathrm{He\,I})$ & $\mathcal{U}(-0.4, 2.2)$ \\
\hline\hline
\colhead{Nebular Parameter} & \colhead{Prior} \\
\hline
$\log U$ & $\mathcal{U}(-4, -1)$\\
$\log n_H$ & $\mathcal{U}(1, 4)$\\
$\mathrm{[O/H]}$ & $\mathcal{U}(-2.2, 0.5)$\\
$\mathrm{[N/O]}$ & $\mathcal{U}(-1, 0.73)$ \\
$\mathrm{[C/O]}$ & $\mathcal{U}(-1, 0.73)$ \\
\enddata
\end{deluxetable}

\par The \texttt{Cloudy} emulator has been trained using a wide range of ionizing spectra, ionization parameters ($U\equiv n_\gamma/n_H$), hydrogen number density ($\log n_H$), oxygen abundance\footnote{\texttt{Cue} fits for gas-phase [O/H], assuming a \emph{total} oxygen abundance of $12+\log(\mathrm{O/H})_\odot=8.93$ \citep{1989Anders}, where $\log(D_\mathrm{O})=-0.22$ dex is depleted onto dust.} (O/H), nitrogen abundance (N/O), and carbon abundance (C/O); these ranges are listed in Table \ref{tab:cuepriors}. All other elemental abundances, relative to hydrogen, are tied to oxygen abundance, assuming a Solar abundance pattern. 
\par The allowed range of ionizing spectra encompasses the spectral shapes that can be expected from a variety of ionizing sources, including young massive stars, post-AGB stars, AGN, and Pop III stars \citep{2025Li}, and the range of ISM parameters is comparable to the ranges typically observed in SFGs and adopted in previous studies of nebular emission \citep{2016Gutkin, 2017Byler}. For our fiducial run of \texttt{Cue}, we adopt the same priors as \cite{2025Li}, which are uniform distributions across the entire parameter space that the neural net has been trained on. 

\begin{figure}
    \centering
    \includegraphics[width=0.95\linewidth]{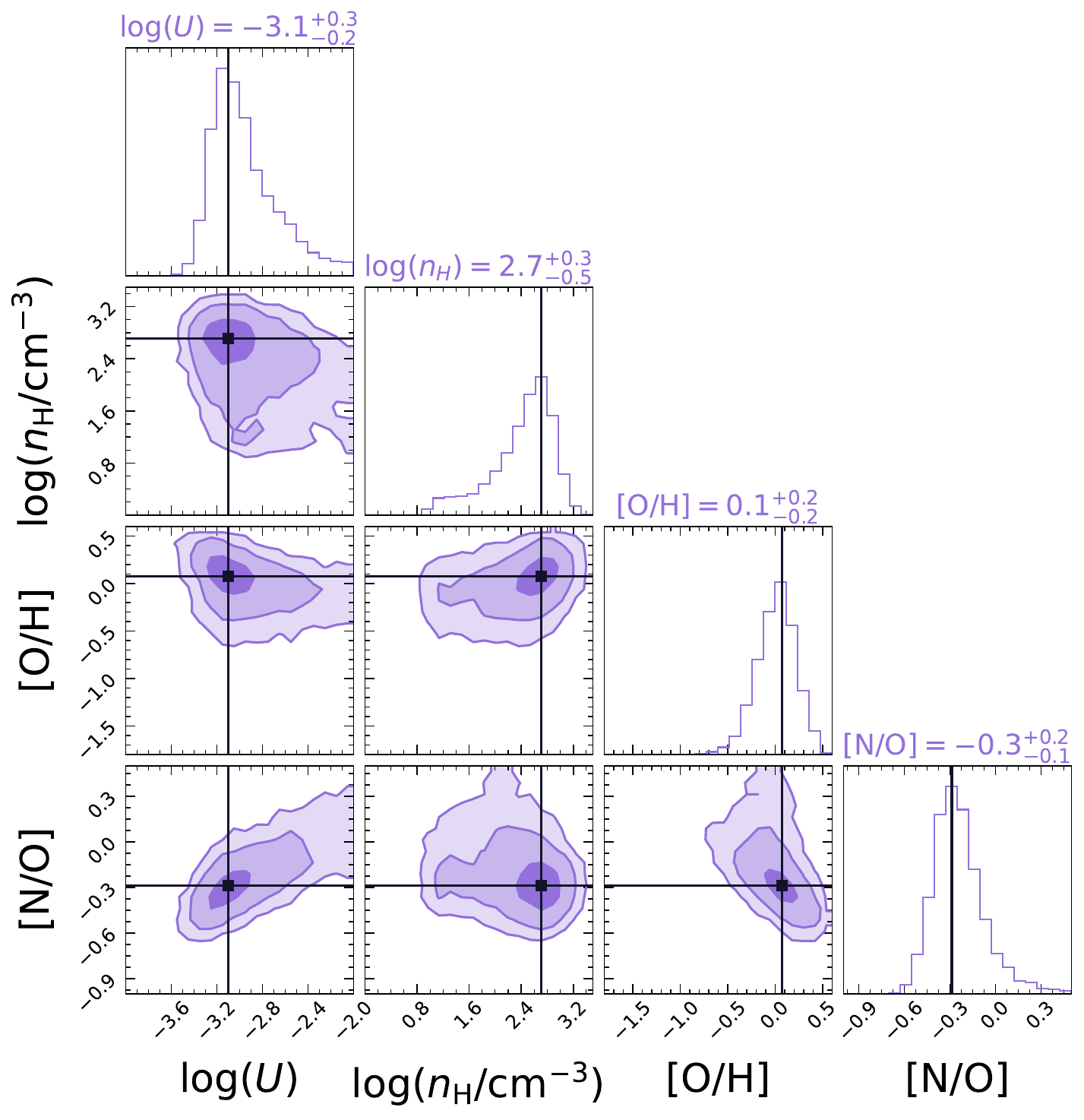}
    \caption{Example of the posterior probability distributions for the nebular conditions inferred by \texttt{Cue} for the CECILIA galaxy Q2343-BX348. Contours show the $16^\mathrm{th},\, 50^\mathrm{th},$ and $84^\mathrm{th}$ percentiles. From left to right or top to bottom these are $\log U$, $\log(n_H)$, [O/H], and [N/O]. Annotations give the MAP (also shown by black lines) and 68\% HDI. }
    \label{fig:corner}
\end{figure}

\begin{figure*}
    \centering
    \includegraphics[width=0.8\linewidth]{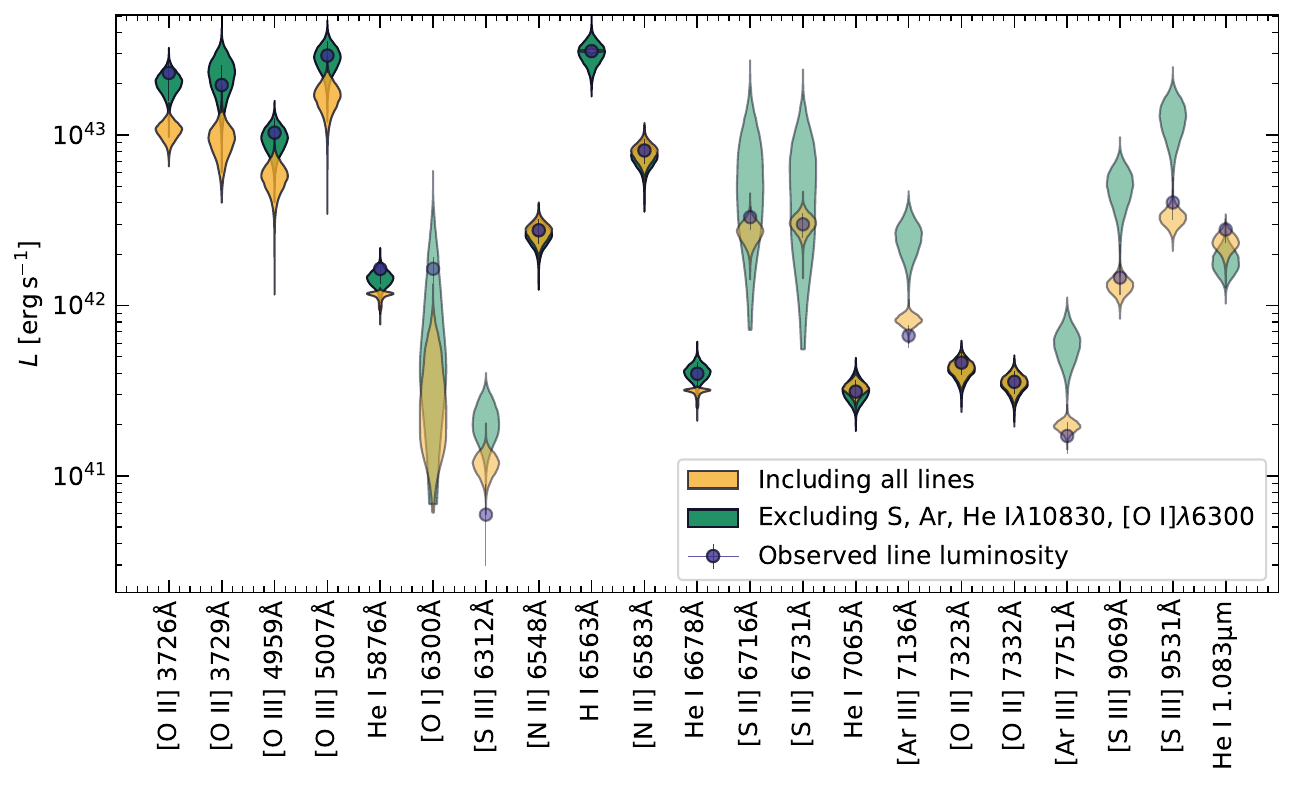}
    \caption{When fitting all available rest-optical to NIR lines, \texttt{Cue} tends to underpredict the luminosity of {[O~III]}$\lambda\lambda 4959,5007$ and {[O~II]}$\lambda\lambda3726,29$ (yellow violin plots), compared to the observed luminosities (purple points with errorbars). These lines are reproduced well in the fiducial run, when sulfur lines, argon lines, {[O~I]}$\lambda6300$, and He~I$\lambda$10830 are excluded (green violins). Lines excluded from the fiducial run are shown by lower opacity violins and points. Modeled and observed luminosities shown here are for Q2343-BX348, which is highlighted in yellow in subsequent figures.}
    \label{fig:cuelines}
\end{figure*}

\par Figure \ref{fig:corner} shows an example posterior probability distribution (CECILIA galaxy Q2343-BX348) for the gas properties that are typically well-constrained; [C/O] is unconstrained because no carbon lines are detected in rest-optical CECILIA or \yp spectra. To systematically identify well-constrained parameters, we adopt the following criteria: firstly, we require that the 68\% highest density interval (HDI) must be separated from the edges of the prior by $\geq1\%$ of the width of the prior range. Additionally, the width of the HDI must be $<$60\% of the prior width. These criteria remove posteriors that are very flat and/or abut the edges of the prior range. We then use a Gaussian mixture model (GMM) to model the shape of the posterior and identify peaks in the distribution. We allow the number of Gaussian components to vary and use the Bayesian information criterion to select the optimal number of components. To remove double-peaked posteriors where neither peak dominates the probability distribution, we check that the minimum probability within the HDI is at the edges of the HDI. Lastly, we check that the maximum a-posteriori (MAP) estimate is not at the edges of the HDI. If the posterior satisfies all of these requirements, we adopt the MAP (taken as the peak of the GMM) as the inferred parameter value, and the bounds of the 68\% HDI as the upper and lower uncertainties. In the case of poorly-constrained posterior probability distributions, we do not report a value. 

\subsection{Applying \texttt{Cue} to CECILIA}\label{sec:nonsolar}
\begin{figure}
    \centering
    \includegraphics[width=1.0\linewidth]{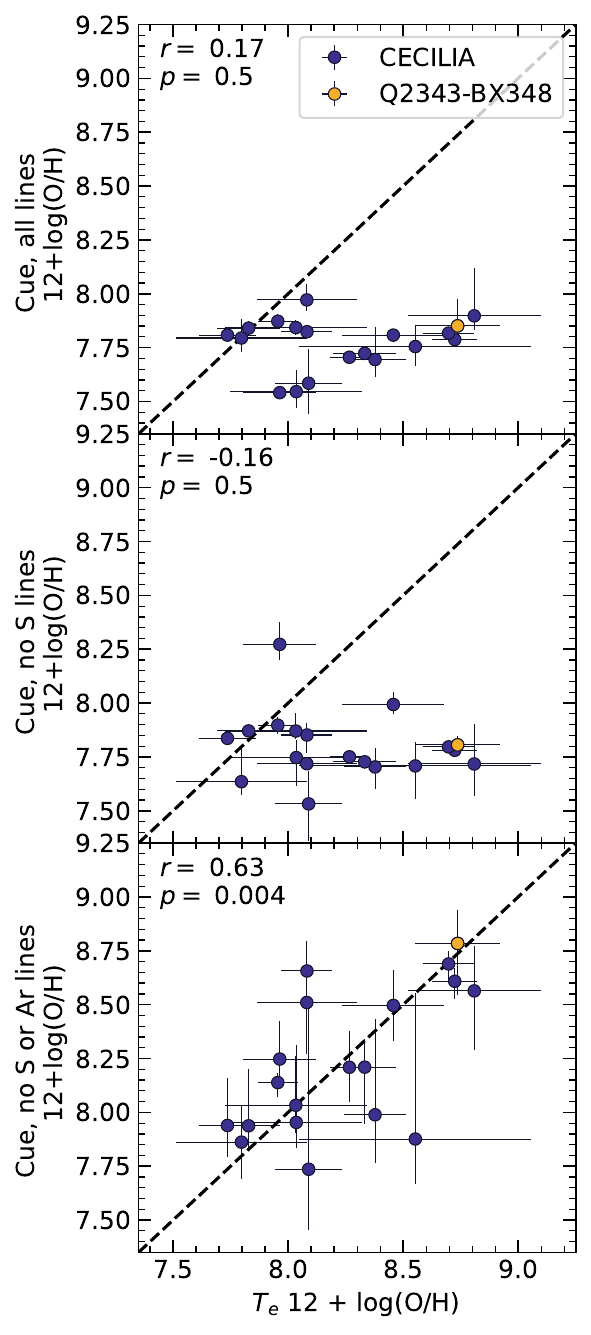}
    \caption{When \texttt{Cue} is run on CECILIA galaxies with all available emission lines (top panel), oxygen abundances (y-axis) cluster around $12+\log(\mathrm{O/H})\sim7.75$, regardless of the $T_e$-based estimate of $12+\log(\mathrm{O/H})$ (x-axis). Excluding just sulfur lines (middle panel) does not resolve this discrepancy. Once sulfur and argon lines are excluded (bottom panel), there is much better agreement between the \ohcue and \ohte, with some scatter around the 1:1 line (black dashed line). Q2343-BX348 (see Figures \ref{fig:corner} and \ref{fig:cuelines}) is highlighted in yellow. Pearson correlation coefficients and $p$-values are given in the corner of each panel.}
    \label{fig:cue_te_comp}
\end{figure}
\par In this section, we assess \texttt{Cue}'s ability to accurately reproduce the observed line luminosities and derived parameters, particularly metallicity, of CECILIA galaxies. We begin by fitting all available emission lines; Figure \ref{fig:cuelines} shows the observed line luminosities (purple points) and posterior distributions (yellow violins) for CECILIA galaxy Q2343-BX348. When all emission lines are included in the fit, the model fails to match the observed nebular {[O~II]} and {[O~III]} line luminosities, despite these being some of the strongest lines in the spectrum, and returns a very narrow range of metallicities, failing to match $T_e$-based measurements of O/H (top panel of Figure \ref{fig:cue_te_comp}). For most galaxies, O/H is significantly underpredicted by \texttt{Cue} when all emission lines are included in the fit. 

\par As \texttt{Cue} uses Solar-scaled abundances for all elements other than oxygen, nitrogen, and carbon, we next assess the impact of including emission lines from other elements on the modeled line luminosities and derived oxygen abundances from \texttt{Cue}. 

\subsubsection{Impact of Non-Solar Abundances}
\par \texttt{Cue} allows O/H, N/O, and C/O to vary independently, but otherwise adopts the Solar abundance pattern from \cite{2000Dopita}. However, recent results from JWST \citep{2024Rogers, 2025aStanton, 2026Foley, 2026Rogers_a} have demonstrated that additional elements, notably sulfur and argon, appear to have non-Solar abundance ratios relative to oxygen. \texttt{Cue} explicitly ties the sulfur and argon abundances to the oxygen abundance via
\begin{equation}
    \log(\mathrm{X/H)=\log(X/O)_\odot+\log(O/H)}
\end{equation}
where the Solar abundance of sulfur or argon relative to oxygen reflects enrichment by both core-collapse and Type Ia supernovae. High-redshift galaxies, however, have enrichment histories that are dominated by CCSNe, resulting in subsolar S/O and Ar/O ratios \citep{2024Rogers, 2024Welch, 2025aStanton, 2026Foley, 2026Rogers_a}. As a result, at fixed O/H, the observed sulfur and argon lines are weaker than what would be predicted under the assumption of Solar S/O and Ar/O ratios. Since the sulfur and argon lines fall in JWST/NIRSpec rather than MOSFIRE, their SNRs are typically higher than that of the nebular oxygen lines, and they are therefore effectively prioritized by the likelihood function. In order to match the observed sulfur and argon line luminosities, a low S/H or Ar/H is needed. Since this can only be changed via O/H, this results in O/H estimates that are biased low. 
\par Figure \ref{fig:cue_te_comp} compares the metallicity predicted by \texttt{Cue} (hereafter \ohcue, y-axis) to the metallicity inferred via the $T_e$ method (hereafter \ohte, x-axis). The top panel shows \ohcue where all measured emission lines are included in the fit (purple points and yellow violins in Figure \ref{fig:cuelines}). Despite the fact that the direct-method abundances suggest that CECILIA galaxies should extend across a broad range in O/H, \texttt{Cue} uniformly infers $12+\log(\mathrm{O/H})\sim7.75$ when sulfur and argon lines are included. 
\par Once sulfur and argon lines are excluded from the fit (third panel in Figure \ref{fig:cue_te_comp}), the agreement between \ohcue and \ohte is significantly improved and there is a strong correlation between the two, with no mean offset between log\ohcue and log\ohte. The middle panel of Figure \ref{fig:cue_te_comp} shows that excluding sulfur lines alone is insufficient to prevent clustering around $12+\log(\mathrm{O/H)_{Cue}}\sim7.75$, implying that \emph{both} sulfur and argon lines contribute to this mismatch. Similar results are seen when only argon lines are excluded. Hence, in our fiducial run of \texttt{Cue} on CECILIA galaxies, we choose to exclude sulfur and argon lines. 
\par When sulfur and argon lines are excluded, \texttt{Cue} predicts systematically higher line fluxes for sulfur and argon lines (green violins in Figure \ref{fig:cuelines}). On average, our fiducial run of \texttt{Cue} predicts {[S~III]}$\lambda\lambda9069,9531$ and {[Ar~III]}$\lambda\lambda7135,7751$ line luminosities that are $\sim3\times$ higher than what is observed. Making the simple assumption that line luminosity scales linearly with elemental abundance (thereby ignoring any possible effects related to the nebular temperature or ionization structure), this implies that the S/O and Ar/O abundances adopted in the model are 0.48 dex too high, and CECILIA galaxies have an average $[\mathrm{S/O]_{Cue}}, [\mathrm{Ar/O]_{Cue}}=-0.48$. \cite{2026Rogers_a} found that CECILIA galaxies have a median $\log(\mathrm{S/O})=-1.78\pm0.21$ and $\log\mathrm{(Ar/O)} = -2.64\pm0.13$. Relative to the \cite{2000Dopita} solar scale, this would imply $[\mathrm{S/O}]_{T_e}=-0.30\pm0.21$, $[\mathrm{Ar/O}]_{T_e}=-0.49\pm0.13$. $[\mathrm{Ar/O}]_{Te}$ is in very good agreement with $[\mathrm{Ar/O]_{Cue}}$, while $[\mathrm{S/O}]_{T_e}$ is slightly higher than $[\mathrm{S/O]_{Cue}}$ (although still within $1\sigma$). The discrepancy may be due to the fact that the $T_e$ calculation of S/H assumes that all {[S~II]}$\lambda\lambda6716,31$ is emitted within the H~II region. However, in our fiducial run, \texttt{Cue} tends to \emph{underpredict} the emission of {[S~II]} relative to the observed line luminosity, implying that some {[S~II]} is emitted outside of the H~II region, causing the $T_e$ method to overestimate S/H by $\sim0.065$ dex \citep[see][and C. von Raesfeld et al., in prep for a more extensive discussion]{2026Rogers_a}. 
\par Modifying \texttt{Cue} to include flexible sulfur and argon abundances is beyond the scope of this paper. However, we advise that sulfur and argon lines should be excluded when non-Solar abundances relative to O/H are suspected but not allowed to vary independently in the photoionization modeling framework. 
\subsubsection{Other Exclusions}
\par There is also evidence to suggest that a substantial fraction of {[O~I]}$\lambda6300$ flux comes from outside of the H~II region (C. von Raesfeld et al., in prep), and therefore, a fully-ionized H~II region model does not accurately describe the line's emission site. Indeed, \texttt{Cue} struggles to match the flux of {[O~I]}$\lambda6300$ (see Figure \ref{fig:cuelines}), systematically underpredicting the observed line luminosity, consistent with additional emission outside of the H~II region. Hence, we choose to exclude it from our fiducial run of \texttt{Cue}.
\par He~I$\lambda10830$ is highly sensitive to density \citep{2014Izotov, 2015Aver, 2026Berg, 2026Skillman} and traces high-density regions in the intermediate- to high-ionization zones. The line becomes especially bright at high densities when collisional excitation begins to populate the $2^3S$ metastable level, from which electrons can be easily excited to the $2^3P$ state and then decay to produce He~I$\lambda10830$. Studies of both local nebulae \citep{2023MendezDelgado_b, 2026Rogers_b} and high-redshift galaxies \citep{2025Martinez, 2025Topping} have shown that the high-ionization zone tends to be denser than the rest of the H~II region, and indeed we find that \texttt{Cue} infers a higher density when He~I$\lambda$10830 is included in the fit. Given that \texttt{Cue} models an H~II region as a sphere of uniform density, we exclude He I$\lambda$10830, and instead allow the density to be determined using the density-sensitive {[O~II]} and {[S~II]} doublets, which trace the intermediate- and low-ionization zones and are used in the determination of $T_e$. When He~I $\lambda10830$ is excluded, \texttt{Cue} is better able to match the observed line luminosities, suggesting that the {[O~II]} and {[S~II]} densities may be more representative of the average density within the H~II region. Models that allow for density variations across different ionization zones may be needed to reproduce the full pattern of observed density-sensitive emission lines  \citep[e.g.,][]{2024Marconi, 2026Moreschini}. However, in the case of models that adopt a single density, we choose to exclude He~I$\lambda10830$.
\par
\begin{figure}
    \centering
    \includegraphics[width=1.0\linewidth]{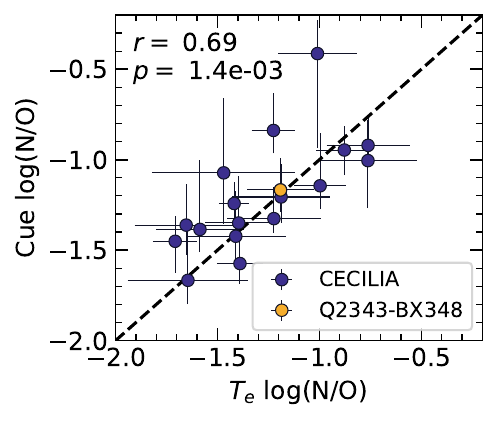}
    \caption{In our fiducial run of \texttt{Cue} on CECILIA galaxies, we recover good agreement between photoionization-model based N/O (y-axis) and $T_e$-based N/O (x-axis). As in Figure \ref{fig:cue_te_comp}, we show the 1:1 line (black dashed line), Q2343-BX348 (yellow), Pearson correlation coefficient, and $p$-value.}
    \label{fig:cecilia_no}
\end{figure}

 Therefore, in our fiducial run of \texttt{Cue} on CECILIA galaxies, we exclude sulfur and argon lines, as well as {[O~I]}$\lambda6300$ and He~I$\lambda10830$. In this case, we find good agreement between \ohcue and \ohte (Figure \ref{fig:cue_te_comp}, bottom panel), as well as $\mathrm{(N/O)_{Cue}}$ and $\mathrm{(N/O)}_{T_e}$ (Figure \ref{fig:cecilia_no}). 
\subsection{Applying \texttt{Cue} to \yp}\label{sec:yp_cue}
\begin{figure}
    \centering
    \includegraphics[width=1.0\linewidth]{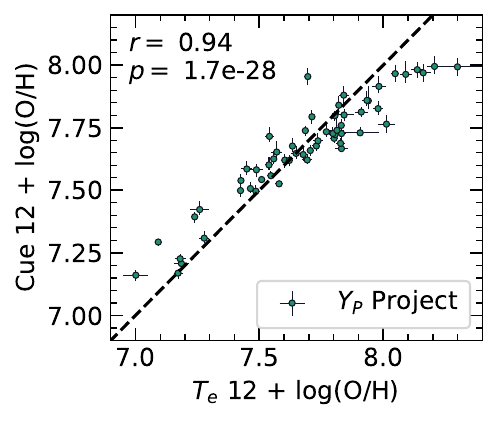}
    \caption{For \yp objects, \ohte (x-axis) and \ohcue (y-axis) are strongly correlated, even when sulfur and argon lines are included in the fit (Pearson correlation coefficient and $p$-value shown). The black dashed line shows the 1:1 line.}
    \label{fig:yp_te_cue}
\end{figure}
\par Unlike CECILIA galaxies, \yp objects have S/O and Ar/O ratios that are relatively consistent with the Solar value assumed in the photoionization model grids. As a result, we can include sulfur and argon lines in our fiducial run while retaining good agreement between \ohcue and \ohte. We do, however, exclude {[O~I]}$\lambda6300$ and He~I$\lambda10830$ for the same reasons outlined in the previous section. 
\par Figure \ref{fig:yp_te_cue} shows good agreement between \ohcue and \ohte. There is no median offset between log\ohcue and log\ohte across the entire sample, but the trend is shallower than the 1:1 line shown, consistent instead with a slope of 0.73. The combination of two effects could create such a trend. Previous photoionization modeling codes have found larger differences between photoionization model and $T_e$ metallicities at low metallicity \citep{2014Perez-Montero}, perhaps consistent with larger temperature fluctuations at low metallicity. This effect could be driving lower-metallicity galaxies towards higher \ohcue. Figure \ref{fig:yp_cecilia_ar_s} shows that \yp objects have slightly subsolar S/O and Ar/O, relative to the adopted Solar scale, but generally have a smaller range in S/O and Ar/O than CECILIA galaxies. Unlike in CECILIA, the difference between log\ohcue and log\ohte is not significantly correlated with S/O or Ar/O, but when S and Ar lines are excluded from the fit, the trend between log\ohcue and log\ohte is slightly steeper (slope $=0.75$), and log\ohcue values are, on average, $\sim0.04$ dex higher. If sulfur and argon lines are affecting the estimation of \ohcue, the effect is quite subtle and secondary to the effect causing higher \ohcue at low O/H; therefore, we choose to keep them in the fiducial run of \texttt{Cue} on \yp objects.
\par As in CECILIA galaxies, we also find good agreement between $\mathrm{(N/O)_{Cue}}$ and $\mathrm{(N/O)}_{T_e}$ for \yp objects ($r=0.86, \, p<10^{-16}$), with no median offset between the two values. The agreement between $\mathrm{(N/O)_{Cue}}$ and $\mathrm{(N/O)}_{T_e}$ is not dependent on N/O.

\section{Measuring Metallicity in Typical Spectra}\label{sec:typicalspec}
\par The overall agreement between \ohcue and \ohte is encouraging, but our fiducial model includes extremely faint emission lines that can be used to make a $T_e$-based measurement of O/H. To use photoionization modeling as a tool to measure O/H in galaxies with shallower spectra, we hope to make equally robust measurements using fewer, brighter emission lines. It is therefore of interest to evaluate how \texttt{Cue}'s performance changes as the available emission lines change and highlight any potential biases introduced. 

\subsection{Nebular Oxygen and Nitrogen Lines}\label{sec:stronglines}
\begin{figure*}
    \centering
    \includegraphics[width=0.8\linewidth]{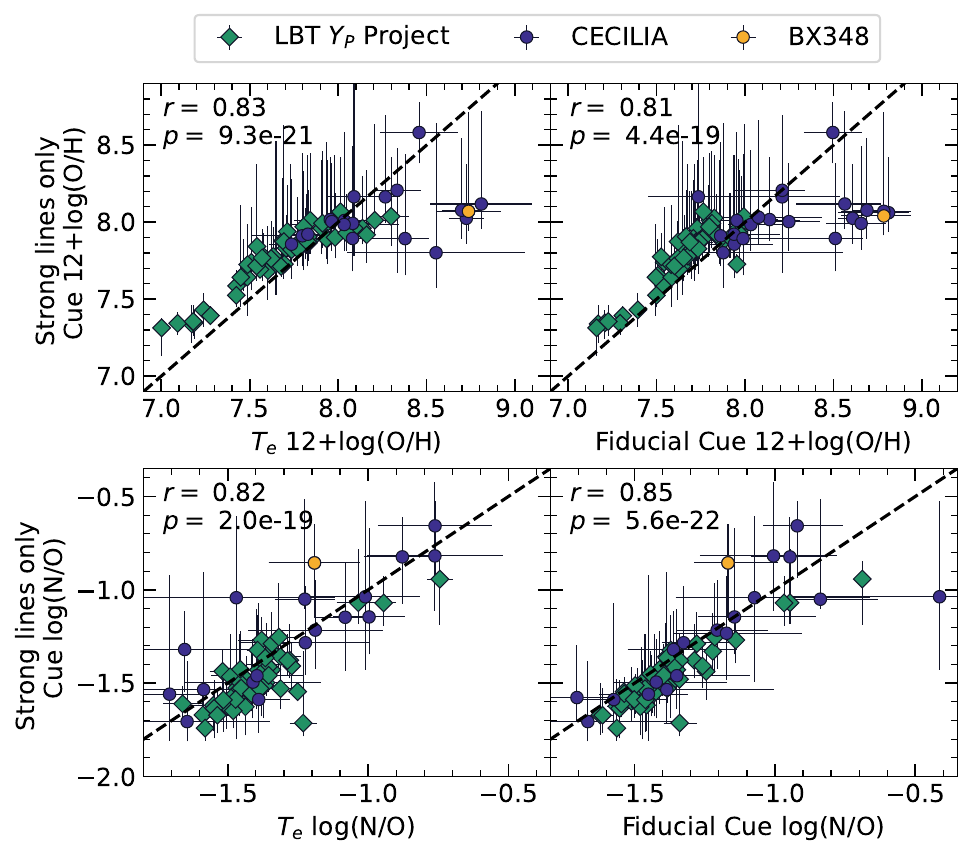}
    \caption{When only strong emission lines are included in the fit, \texttt{Cue} cannot accurately infer O/H, but can infer N/O. \emph{Top left:} \ohstrong vs. \ohte. \emph{Top right:} \ohstrong vs. \ohfull. \emph{Bottom left:} $\mathrm{(N/O)_{Cue,strong}}$ vs. $\mathrm{(N/O)}_{T_e}$. \emph{Bottom right:} $\mathrm{(N/O)_{Cue,strong}}$ vs. $\mathrm{(N/O)_{fid}}$. In all panels, the black dashed line shows the 1:1 line, CECILIA galaxies are shown by purple circles, Q2343-BX348 is highlighted in yellow, \yp objects are shown by green diamonds, and the Pearson correlation coefficient and $p$-value are given.}
    \label{fig:cecilia_strong}
\end{figure*}
\par Strong-line ratios such as O3 and R23 are double-valued as a function of metallicity, resulting in degenerate high- and low-metallicity solutions (see Figure \ref{fig:yp_cecilia_r3}). Past work has shown that photoionization modeling of strong emission lines alone cannot break this degeneracy without assuming some $U$-O/H or N/O-O/H correlation \citep[e.g., ][]{2014Perez-Montero, 2016ValeAsari, 2018Strom}. Hence we begin by assessing \texttt{Cue}'s performance when only strong nebular lines are included in the fit. We run \texttt{Cue} on CECILIA and \yp objects using only {[O~II]}$\lambda\lambda3726,29$, {[O~III]}$\lambda\lambda4959,5007$, H$\alpha$, and {[N~II]}$\lambda\lambda6548,85$ (line list 1 in Table \ref{tab:line_lists}, note that while H$\beta$ is not explicitly included in the fit, it is used in dust-correction). Figure \ref{fig:cecilia_strong} shows the recovered O/H and N/O (hereafter referred to as \ohstrong and (N/O)$_\mathrm{Cue, strong}$) in comparison to both $T_e$-based estimates and estimates from the fiducial runs of \texttt{Cue}. 
\par Overall, there is a significant correlation between \ohstrong and both \ohte and \ohcue when considering both CECILIA and \yp. However, there is an offset towards higher log\ohstrong at fixed log\ohte (and log\ohcue), \ohstrong tends to have large and asymmetric uncertainties, and the correlation breaks down at higher O/H. In fact, the correlation becomes statistically insignificant ($p>0.05$) if objects with $12+\log\ohte \leq 7.83$ are excluded. There is also no statistically significant correlation when considering CECILIA galaxies alone. Therefore, we can surmise that the Spearman $r$ and $p$-values shown in Figure \ref{fig:cecilia_strong} are primarily driven by better agreement at low O/H.  
\par The observed trend suggests that \texttt{Cue}, like other photoionization modeling frameworks, cannot reliably break the degeneracy between low- and high-metallicity solutions. While we see some agreement between \ohcue and \ohte on the low-metallicity branch and near the turnover, this agreement breaks down on the high-metallicity branch, where the sampling algorithm returns an intermediate solution between the high- and low-O/H branches, causing all objects to cluster around the turnover metallicity ($12+\log(\mathrm{O/H})\sim8.1$, also coincident with the median of the prior).  Therefore, we conclude that \texttt{Cue} cannot reliably estimate O/H when using strong nebular emission lines alone.
\par In contrast, we find that N/O can be accurately measured with strong emission lines alone. There is good agreement between $\mathrm{(N/O)_{Cue,strong}}$)and both (N/O)$_{T_e}$ and $\mathrm{(N/O)_{Cue}}$ from the fiducial run. Such agreement is consistent across the full range of N/O and the correlation remains significant when only considering the CECILIA sample. 
\par Obtaining a direct measurement of $T_e$({[N~II]}), and therefore N/H and N/O, is difficult due to the faintness of the {[N~II]}$\lambda5755$ line. However, because N/O can be accurately constrained with strong nebular lines, we can circumvent the need for very faint auroral lines. N/H, on the other hand, cannot be reliably inferred with strong nebular lines due to the poor recovery of O/H.  Additionally, the agreement between $\mathrm{(N/O)_{Cue,strong}}$ and (N/O)$_{T_e}$ suggests that strong-line diagnostics combining nebular oxygen and nitrogen emission lines, such as O3N2, are likely primarily sensitive to N/O, and hence using these diagnostics to estimate O/H is highly sensitive to the O/H-N/O correlation in the calibration sample. This may be particularly problematic at high-$z$ where anomalously high N/O is sometimes observed. 

\subsection{Adding [Ne III]$\lambda3869$}\label{sec:neiii}
\begin{figure*}
    \centering
    \includegraphics[width=0.95\linewidth]{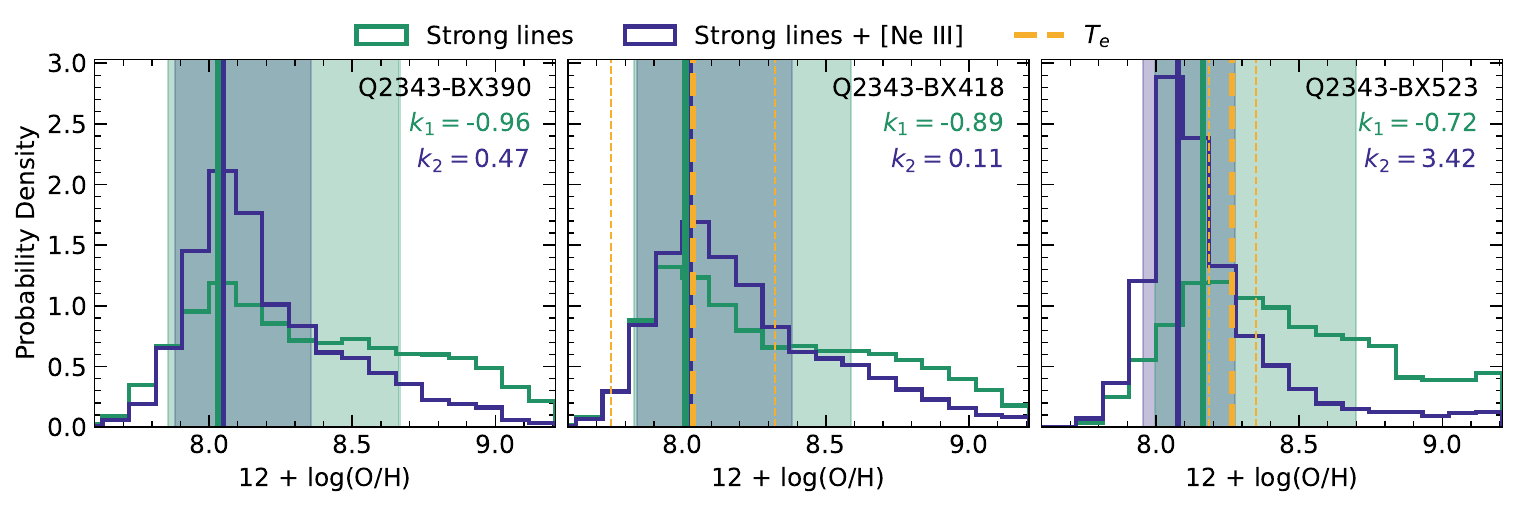}
    \caption{Strong lines + {[Ne~III]} (purple histograms and shaded regions) can result in more sharply peaked posterior probability distributions and narrower 68\% HDIs (shaded regions) than strong lines alone (green histograms and shaded regions). To quantify the change in shape, we report the Fisher kurtosis of each posterior. The effect on the MAP (solid vertical lines), as well as the agreement between \ohcue and \ohte (yellow dashed line) is less clear from this small sample.}
    \label{fig:neiii_posteriors}
\end{figure*}
\par Semi-strong emission lines have been shown to be highly effective in breaking the degeneracy in O/H \citep{2016ValeAsari}. However, {[Ar~III]}$\lambda7135,7751$ and {[S~III]}$\lambda\lambda 9069,9531$ significantly bias the results of photoionization modeling with \texttt{Cue} in high-redshift galaxies due to subsolar S/O and Ar/O (see Section \ref{sec:nonsolar}). {[Ne~III]}$\lambda3869$ may be a more promising candidate as neon is, like oxygen, a typical $\alpha$-element with no contribution from Type Ia SNe \citep{2020Kobayashi}. As a result, Ne/O is expected to be reasonably constant, regardless of a galaxy's SFH, and indeed high-$z$ galaxies tend to show Solar Ne/O \citep{2022ArellanoCordova, 2024MarquesChaves} but subsolar S/O and Ar/O \citep{2024Rogers, 2024Welch, 2025aStanton, 2026Foley, 2026Rogers_a}. While there is some evidence for enhanced Ne/O at high O/H in local galaxies \citep{2023MirandaPerez, 2024Arellano-Cordova} and subsolar Ne/O at $6<z<10$ \citep[][potentially due to the presence of very massive stars, \citealt{2024Watanabe}]{2023Isobe}, Ne/O appears to be consistent with Solar in galaxies similar to CECILIA galaxies. Therefore, including neon emission lines in our photoionization model fits should not bias O/H estimates due to the assumption of Solar Ne/O. 
\par {[Ne~III]}$\lambda3869$ is not accessible in G235M for CECILIA galaxies and is measured with a SNR $>$ 3 in only 6 MOSFIRE spectra. However, this line is much more readily observed with JWST (with $\sim3500$ {[Ne~III]}$\lambda3869$ detections at SNR $>3$ in the DAWN JWST Archive\footnote{\url{https://dawn-cph.github.io/dja/blog/2025/05/01/nirspec-merged-table-v4/}}). Unfortunately, most of the CECILIA galaxies with {[Ne~III]}$\lambda3869$ are also near the turnover metallicity, making it difficult to systematically assess whether the line is helpful in breaking the degeneracy between high- and low-metallicity solutions. 
\par We show here three promising examples where adding {[Ne~III]} significantly changes the shape of the posterior probability distribution (Figure \ref{fig:neiii_posteriors}). In all three cases, adding {[Ne~III]} results in a more sharply peaked distribution with a weaker wing towards higher O/H (purple histograms). The change in shape is reflected in the Fisher kurtoses of the posteriors; in all three cases, the kurtosis increases from a negative value (platykurtic, or more broadly peaked) to a positive value (leptokurtic, or more sharply peaked). The 68\% HDI is also, on average, 0.31 dex narrower, suggesting that {[Ne~III]} can significantly improve the precision of \ohcue. 
\par Though it may improve precision, it is unclear whether adding {[Ne~III]} improves the accuracy of \ohcue. \ohte is not measured for Q2343-BX390 (left panel), and the two MAP values are very similar. The MAP values are also very similar for Q2343-BX418 (middle panel), and both coincide closely with \ohte. The MAP value is slightly lower when {[Ne~III]} is added to Q2343-BX523 (right panel) bringing it further from the \ohte estimate (although still within $1\sigma$ of \ohte). A similar trend is seen in \yp objects; adding {[Ne~III]} results in smaller, more symmetric uncertainties, but does not significantly improve the agreement between O/H found using strong lines and {[Ne~III]} and \ohte. However, we have shown in Section \ref{sec:stronglines} that the agreement between \ohstrong and \ohte is highly dependent on O/H, so a larger sample of galaxies at a range of O/H is needed to fully assess the utility of {[Ne~III]} in photoionization modeling. 

\subsection{Varying Spectral Depth}\label{sec:yp_1234}
\par In objects with Solar S/O and Ar/O, we can use ``semi-strong" sulfur and argon lines without significantly biasing our estimate of \ohcue, giving us a much larger list of lines to explore. 
\par To assess the impact of including increasingly faint emission lines on \texttt{Cue}'s performance, we define four different line lists, each including subsequently fainter lines, and compare the results of photoionization modeling using each list. In the case of fixed ratio doublets, we use the average luminosity of the brighter line in the pair to determine which list both lines fall into. The four line lists are listed in Table \ref{tab:line_lists}.

\begin{deluxetable*}{c|c|>{\raggedright\arraybackslash}p{8cm}|c|c|c}
\tablecaption{Summary of the different line lists given to \texttt{Cue} to estimate O/H in \yp objects, median offset between \ohsubset and \ohfull ($\Delta$log(O/H)), median uncertainty on \ohsubset ($\sigma\ohsubset$), and RMS values of $\Delta$log(O/H) \label{tab:line_lists}}
\tablehead{
\colhead{} & \colhead{List Name} & \colhead{Lines Included\tablenotemark{a}} & \colhead{$\Delta\log(\mathrm{O/H})$} & \colhead{$\sigma\ohsubset$} & \colhead{RMS($\Delta\log(\mathrm{O/H})$)}
}
\startdata
\multirow{4}{*}{\rotatebox[origin=c]{90}{\makebox[3cm][c]{\parbox{3cm}{\centering Changing depth}}}} & 1& {[O~II]}$\lambda\lambda3726,29$, {[O~III]}$\lambda\lambda4959,5007$, {[N~II]}$\lambda\lambda 6548,83$, and H$\alpha$ & 0.13 & $^{+0.29}_{-0.13}$ & 0.14\\
\cline{2-6}
 & 2 & List 1 + {[S~III]}$\lambda\lambda9069,9531$, {[S~II]}$\lambda\lambda6716,6731$, He~I$\lambda5876$, and {[Ne~III]}$\lambda 3867$ & 0.01 & $^{+0.03}_{-0.03}$ & 0.06 \\
\cline{2-6}
 & 3 & List 2 + {[Ar~III]}$\lambda\lambda 7135, 7751$ and He~I$\lambda$7065 & 0.01 & $^{+0.03}_{-0.03}$ & 0.06\\
\cline{2-6}
 & 4 & List 3 + {[O~III]}$\lambda4363$, He~I$\lambda6680$, {[O~II]}$\lambda7323,32$, He~II$\lambda4686$, and {[S~III]}$\lambda 6312$ & 0.01 & $^{+0.03}_{-0.02}$ & 0.03\\
\hline\hline
\multirow{4}{*}{\rotatebox[origin=c]{90}{\makebox[5cm][c]{\parbox{4.5cm}{\centering Changing wavelength coverage}}}}
& 2(a) &  Same as list 2 ($3726\ang \leq \lambda_\mathrm{rest}\leq9531\ang$): {[O~II]}$\lambda\lambda$3726,29, {[Ne~III]}$\lambda 3867$, {[O~III]}$\lambda\lambda 4959,5007$, He~I$\lambda$5876, {[N~II]}$\lambda\lambda 6548,6583$, H$\alpha$, {[S~II]}$6716,6731$, and {[S~III]}$\lambda\lambda9069,9531$ & 0.01 & $^{+0.03}_{-0.03}$ & 0.06 \\
\cline{2-6}
 & 2(b) & List 2 lines bluewards of [S II] ($3726\ang \leq \lambda_\mathrm{rest}\leq6731\ang$): {[O~II]}$\lambda\lambda$3726,29, {[Ne~III]}$\lambda 3867$, {[O~III]}$\lambda\lambda 4959,5007$, He~I$\lambda$5876, {[N~II]}$\lambda\lambda 6548,6583$, H$\alpha$, and {[S~II]}$6716,6731$ & 0.10 & $^{+0.10}_{-0.05}$ & 0.14 \\
\cline{2-6}
 & 2(c) & List 2 lines bluewards of [N II] ($3726\ang \leq \lambda_\mathrm{rest}\leq6583\ang$): {[O~II]}$\lambda\lambda$3726,29, {[Ne~III]}$\lambda 3867$, {[O~III]}$\lambda\lambda 4959,5007$, He~I$\lambda$5876, {[N~II]}$\lambda\lambda 6548,6583$, and H$\alpha$ & 0.10 & $^{+0.13}_{-0.09}$ & 0.14\\
\cline{2-6}
 & 2(d) & List 2 lines redwards of H$\beta$ ($4861\ang \leq \lambda_\mathrm{rest}\leq9531\ang$): {[O~III]}$\lambda\lambda 4959,5007$, He~I$\lambda$5876, {[N~II]}$\lambda\lambda 6548,6583$, H$\alpha$, {[S~II]}$6716,6731$, and {[S~III]}$\lambda\lambda9069,9531$ & 0.02 & $^{+0.06}_{-0.04}$ & 0.06\\
\enddata
\tablenotetext{a}{Note that while H$\beta$ is not explicitly included in the fit, the line is needed in order to correct line luminosities for dust-reddening.}
\end{deluxetable*}
\begin{figure*}
    \centering
    \includegraphics[width=0.8\linewidth]{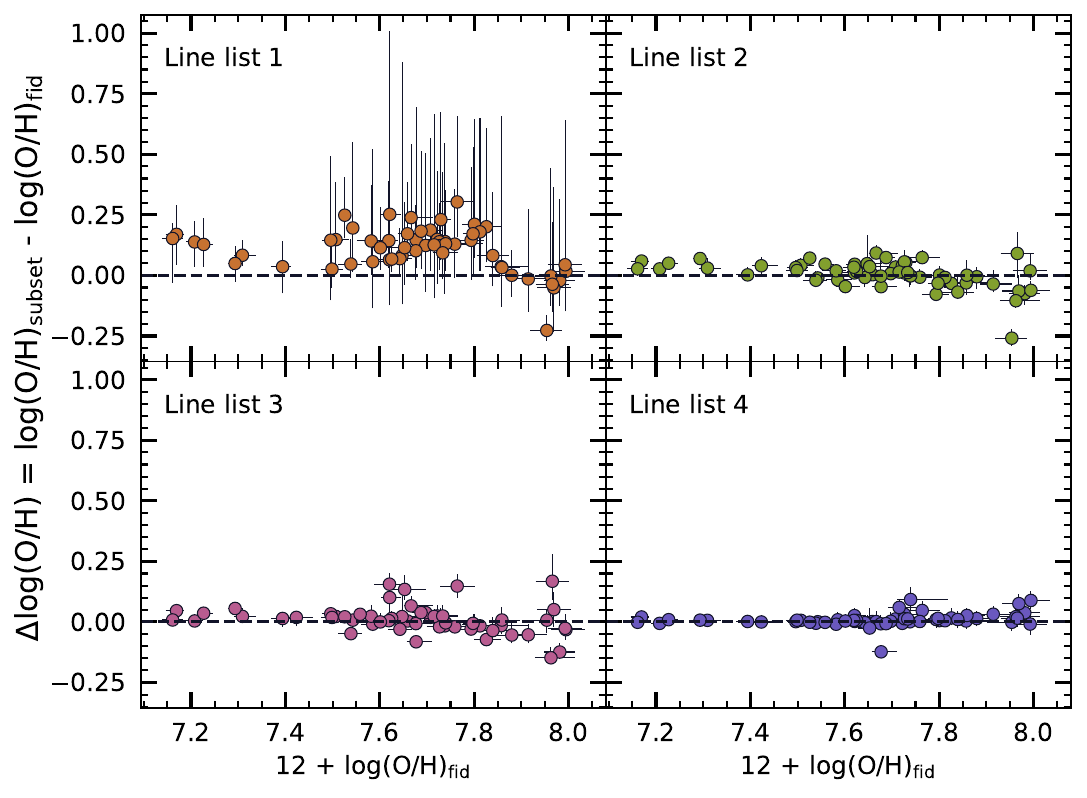}
    \caption{As fainter emission lines are included in the fit of \yp objects, \ohcue is recovered with higher fidelity and less uncertainty.  Line list 2 results in reasonably good recovery of \ohfull. Each panel plots the difference between log\ohsubset and log\ohfull against \ohfull. The lines included in each subset, median offsets, median uncertainties, and RMS offsets are listed in Table \ref{tab:line_lists}. }
    \label{fig:yp_1234}
\end{figure*}
\par Figure \ref{fig:yp_1234} shows the results of fitting \yp objects with each line list. We compare the \ohcue value inferred using a given line subset, referred to as $\mathrm{(O/H)_{subset}}$, to \ohfull (see Figure \ref{fig:line_list_summary} for the full set of lines included in the fiducial run). Table \ref{tab:line_lists} lists the emission lines included in each subset, the median difference between log$\mathrm{(O/H)_{subset}}$ and log\ohfull ($\Delta\log(\mathrm{O/H})$), uncertainty on $\mathrm{(O/H)_{subset}}$, and root-mean-square of $\Delta\log(\mathrm{O/H})$. 
\par As is discussed in Section \ref{sec:stronglines}, when only strong lines are included (line list 1), $\mathrm{(O/H)_{subset}}$ tends to be higher than \ohfull, with large, lopsided uncertainties, reflecting weakly double-peaked posteriors. Adding the semi-strong {[Ne~III]}$\lambda3867$, He~I$\lambda5876$, {[S~II]}$\lambda\lambda6716,6731$, and {[S~III]}$\lambda\lambda9069,9531$ (line list 2) breaks the degeneracy between low- and high-O/H solutions. As a result, the typical uncertainty on O/H is much smaller (0.03 dex) and more symmetric. The offset towards larger \ohsubset is also eliminated. Adding the weaker {[Ar~III]} lines (line list 3) does not significantly improve the precision or accuracy of the fit, while adding auroral lines (line list 4) predictably leads to the best agreement between \ohsubset and \ohfull. Therefore, the lines added in line list 2 represent the largest gain in fidelity relative to strong emission lines alone. Promisingly, these lines are typically much stronger than auroral emission lines, and therefore require shorter exposure times. For high-redshift galaxies, these lines can be observed from both the ground and space. However, as noted in Section \ref{sec:nonsolar}, these lines should not be used in conjunction with Solar-scaled photoionization models if non-Solar abundance patterns are suspected. 

\subsection{Varying Wavelength Coverage}\label{sec:yp_abcd}
\par
\begin{figure*}
    \centering
    \includegraphics[width=0.8\linewidth]{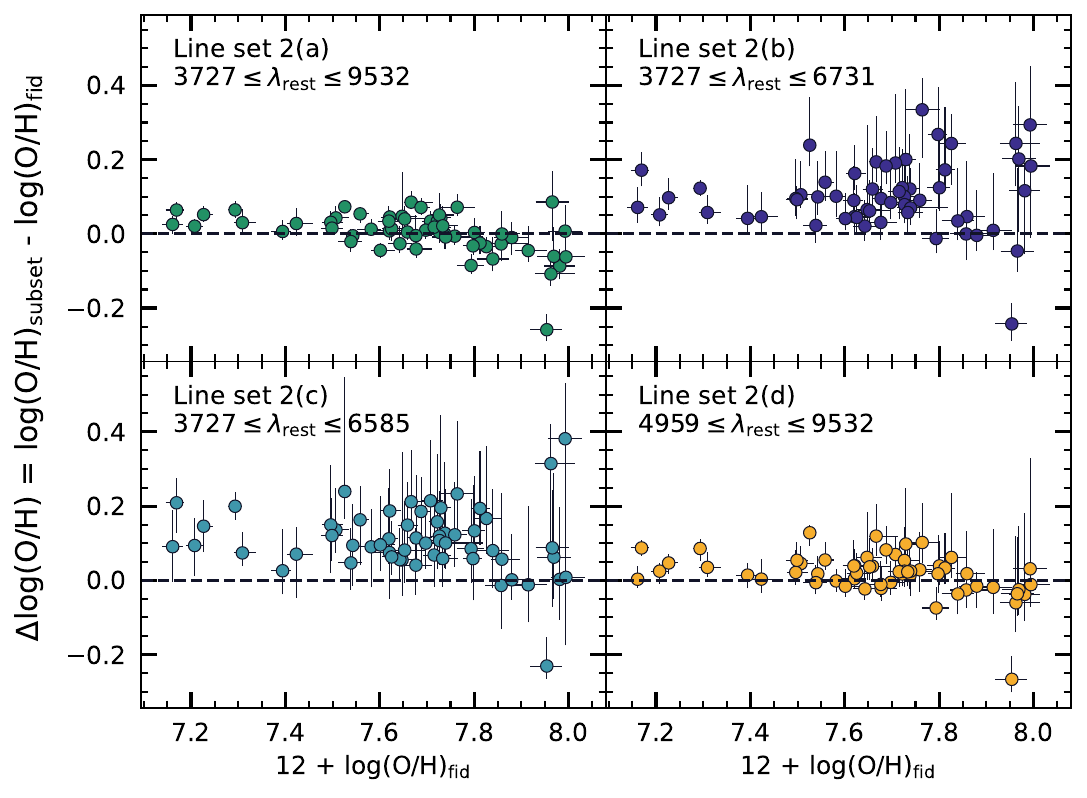}
    \caption{{[S~III]}$\lambda\lambda$9069,9531, which traces the intermediate-ionization zone, is crucial in producing accurate estimates of \ohcue for \yp objects, resulting in good agreement between \ohsubset and \ohfull for line lists 2(a) and 2(d). As in Figure \ref{fig:yp_1234}, each panel plots the difference between log\ohsubset and log\ohfull versus \ohfull. Summary statistics are given in Table \ref{tab:line_lists}.}
    \label{fig:yp_abcd}
\end{figure*}

 Nonetheless, the emission lines included in line list 2 span a large range in wavelength ($3726\ang \leq\lambda_\mathrm{rest}\leq 9531\ang$) and require observations in multiple filters or instrument configurations. To identify the minimal wavelength coverage needed to make robust estimates of \ohcue, we define line lists 2(a) through 2(d) (see Table \ref{tab:line_lists}), which make wavelength-dependent cuts to the lines included in line list 2. Figure \ref{fig:yp_abcd} shows the results of fitting with each list. 
\par Line list 2(b) (top right panel) excludes the reddest emission lines ({[S~III]}$\lambda\lambda9069,9531$). This both increases the median uncertainty on $\mathrm{(O/H)_{subset}}$ (from $^{+0.03}_{-0.03}$ dex to $^{+0.10}_{-0.05}$ dex) and introduces a slight offset (0.10 dex) with respect to \ohfull. Excluding {[S~III]} and {[S~II]} (line list 2(c), bottom left panel) further increases the median uncertainty to $^{+0.13}_{-0.09}$ dex, consistent with what is seen when {[Ne~III]}$\lambda3867$ is added to the strong nebular lines (Section \ref{sec:neiii}). 
\par When the bluest emission lines ({[O~II]} and {[Ne~III]}) are excluded but {[S~II]} and {[S~III]} are included (line list 2(d), bottom right), the offset with respect to \ohfull is small and comparable to the offset seen with line list 2(a) (0.02 dex) and the precision is improved relative to what is found with lists 2(b) and 2(c). Additionally, the RMS of $\Delta$log(O/H) for line list 2(d) is comparable to that of set 2(a), suggesting the systematic uncertainty on \ohcue does not change significantly when {[O~II]} and {[Ne~III]} are excluded. Of the semi-strong emission lines included in line list 2(d), we find that {[S~III]} has the largest effect on the recovered value of \ohcue, and excluding these lines reintroduces the offset to higher log\ohcue seen in line lists 2(b) and 2(c). Notably, {[S~III]} is the only line in the superset (line list 2(a)) that exclusively probes the intermediate ionization zone, suggesting that \ohcue is most accurately recovered when fitting lines originating in all three ionization zones. This may explain why {[S~III]} appears to have a stronger effect in constraining \ohcue as compared to {[Ne~III]}, which traces the high-ionization zone. 
\par Therefore, we find that \texttt{Cue} can accurately recover O/H with just nebular {[O~III]}, {[N~II]}, {[S~II]}, and {[S~III]} in relatively metal-poor ($12+\log\ohte\lesssim 8.3$) galaxies with S/O roughly consistent with Solar. It is possible that {[O~II]} may be more important in modeling more metal-rich systems, which may have softer ionizing spectra and fractionally more oxygen in the singly ionized state \citep{1988Vilchez}, but further work is needed to assess this. \texttt{Cue} can therefore be used to make accurate estimates of O/H in samples with detections of lines $\sim 10\times$ fainter than H$\alpha$, provided that the photoionization model grids assume an appropriate S/O.

\section{Conclusions and Recommendations}\label{sec:recs}
\par In this letter, we have used $z\sim2.3$ galaxies from the CECILIA survey and local galaxies from the \yp to assess the precision and accuracy of using photoionization models to measure metallicity in different circumstances. 
\begin{itemize}
    \item We find that \texttt{Cue} provides a promising framework for measuring O/H while also accounting for the effects of ionization parameter, density, and the ionizing spectrum on the emission line spectrum of a galaxy. In our fiducial run of \texttt{Cue}, we recover good agreement between \ohte and \ohcue across a wide range of O/H (Section \ref{sec:phot_model}).
    \item Due to subsolar S/O and Ar/O ratios in SFGs at $z>2$, we exclude sulfur and argon emission lines when modeling the CECILIA galaxies (Section \ref{sec:nonsolar}), but find that these lines can be included when modeling local galaxies, which are more consistent with the Solar abundance pattern, without biasing the estimation of O/H (Section \ref{sec:yp_cue}).
    \item In local galaxies with Solar S/O and Ar/O abundances, we find that \ohcue can be recovered using only the nebular {[O~III]}, {[N~II]}, {[S~II]}, and {[S~III]} lines (Sections \ref{sec:yp_1234} and \ref{sec:yp_abcd}). This ensemble of lines traces the low-, intermediate-, and high-ionization zones within H II regions. 
\end{itemize}
\par The tests presented in this letter have implications for the application of \texttt{Cue} and other photoionization modeling frameworks to both local and high-redshift galaxies, as well as for future developments in photoionization modeling. 
\par \emph{Recommendations for high-redshift galaxies:} \texttt{Cue} and other photoionization models that assume Solar scaling of elements such as sulfur and argon with respect to oxygen should be used with caution at high redshift. When measuring O/H, emission lines from other elements should only be used if the element has the same enrichment pathway as oxygen (i.e., no Type Ia contribution, for example, neon) or if the abundance of that element is allowed to vary within the photoionization model (carbon and nitrogen, in the case of \texttt{Cue}). Otherwise, the resulting O/H estimates will be severely biased by the abundance ratios of other elements (Section \ref{sec:nonsolar}, Figure \ref{fig:cue_te_comp}). 
\par Due to the double-valued behavior of strong nebular emission lines, photoionization models cannot be used to uniquely infer the metallicity of a galaxy when only strong emission lines are available (Section \ref{sec:stronglines}, Figure \ref{fig:cecilia_strong}). {[Ne~III]}$\lambda$3869 is the main semi-strong $\alpha$-element emission line that could serve to break the degeneracy between low- and high-metallicity solutions. While preliminary testing on the CECILIA sample suggests the line could be useful (Section \ref{sec:neiii}, Figure \ref{fig:neiii_posteriors}), a larger sample of high-redshift galaxies with {[Ne~III]} detections is needed to fully assess the line's utility. The issue of mismatched abundance patterns currently limits the applicability of \texttt{Cue} at high redshift to samples with {[Ne~III]} detections or auroral emission lines.

\par \emph{Recommendations for local galaxies:} the \yp objects have S/O and Ar/O abundances that are more well-matched by the model grids that \texttt{Cue} is trained on. We show that, in this case, semi-strong sulfur and neon lines can be leveraged to measure oxygen abundances with relatively high fidelity (Section \ref{sec:yp_1234}, Figure \ref{fig:yp_1234}). In particular, we find that {[S~III]}$\lambda\lambda$9069,9531 can significantly improve the precision of metallicity estimates (Section \ref{sec:yp_abcd}, Figure \ref{fig:yp_abcd}), suggesting that relatively shallow follow-up in the red-optical/NIR could be especially valuable for studies of local galaxy metallicities. The brighter {[S~III]} line can be of comparable strength to or even stronger than {[N~II]}$\lambda6583$ but almost always requires multi-band spectroscopy due to the large wavelength separation between it and other commonly observed emission lines. Studies aiming to measure metallicity using \texttt{Cue} should therefore prioritize wide wavelength coverage over deeper spectra in a single band. 

\par \emph{Recommendations for future photoionization modeling efforts:} The results of this paper suggest that accounting for non-Solar abundances is key to interpreting emission lines from elements that are enriched via multiple nucleosynthetic channels. This is of particular relevance to high-redshift galaxies, which tend to be dominated by CCSN enrichment, where Solar-scaled models cannot reproduce observed emission line patterns (Section \ref{sec:nonsolar}). Therefore, photoionization model grids with subsolar S/O and Ar/O are necessary to model the emission line spectra of high-redshift galaxies. 
\par Flexibly modeling galaxies with a range of SFHs will require photoionization model grids spanning a range of elemental abundances relative to oxygen. The abundance of each relevant element could be individually fit, or multiple element abundances could be tied together by a single parameter, such as the degree of Type Ia SN enrichment or a single, representative abundance, assuming some correspondence between different elemental abundances. Further training of the neural net emulator on model grids with varying S/O and Ar/O is beyond the scope of this paper but would significantly expand \texttt{Cue}'s applicability to high-redshift galaxies. 
\par Alternatively, lines from elements affected by non-Solar abundance patterns can still be used as diagnostics of ionization and spectral hardness. Ratios of ions in different ionization zones (e.g., {[S~III]}$\lambda\lambda9069,9531$/{[S~II]}$\lambda\lambda 6716,6731$ or [Ar IV]$\lambda\lambda 4711,4740$/{[Ar~III]}$\lambda\lambda7135,7751$) are highly sensitive to ionization parameter and have secondary sensitivities to spectral hardness \citep{2021Berg}, but they are, to first order, insensitive to elemental abundance (C. von Raesfeld et al., in prep). Modifying likelihood functions to fit {[S~III]}/{[S~II]} or [Ar IV]/{[Ar~III]} ratios, rather than absolute line luminosity, would add information to photoionization modeling without introducing a bias in gas-phase metallicity. 

\begin{acknowledgements}
\par N.K.C. would like to thank Alice Shapley for her insightful comments on the analysis presented in this paper. N.K.C. is supported by the National Science Foundation Graduate Research Fellowship Program under Grant No. 2025381248. N.K.C. acknowledges use of Claude (Anthropic) as a tool in debugging, particularly in figure and \LaTeX\xspace formatting; all scientific content, analysis, and conclusions are the authors'. 
\par G.C.R., R.F.T, and A.L.S. acknowledge partial support from the JWST-GO-02593.006-A, JWST-GO-02593.008-A, and JWST-GO-02593.004-A grants, respectively. N.S.J.R. was also supported by JWST-GO-02593.008-A. These funds were provided by NASA through a grant from the Space Telescope Science Institute, which is operated by the Association of Universities for Research in Astronomy, Inc., under NASA contract NAS503127. G.C.R. acknowledges the support from Grant 63667 from the John Templeton Foundation. Any opinions, findings, and conclusions or recommendations expressed in this material are those of the authors and do not necessarily reflect the views of the John Templeton Foundation. R.F.T. also acknowledges support from the Pittsburgh Foundation (grant ID UN2021-121482) and the Research Corporation for Scientific Advancement (Cottrell Scholar Award, grant ID 28289). A.L.S. is also supported by the David and Lucile Packard Foundation (Packard Fellowship, grant 2024-77399) and the National Science Foundation (grant number AST-2406780). Y.L. is supported by the National Science Foundation (grant number AST-2406780) and a Research Corporation for Scientific Advancement (RCSA) Scialog Award (grant number SA-LSST-2024-094a). Z.Z. acknowledges the financial support from the CIERA Postdoctoral Fellowship.
\par This work is based on observations made with NASA/ESA/CSA JWST, associated with PID 2593, which can be accessed via \href{https://archive.stsci.edu/doi/resolve/resolve.html?doi=10.17909/x66z-p144}{doi:10.17909/x66z-p144}, the W. M. Keck Observatory, and the Large Binocular Telescope. The LBT is an international collaboration among institutions in the United States, Italy and Germany. LBT Corporation Members are: The University of Arizona on behalf of the Arizona Board of Regents; Istituto Nazionale di Astrofisica, Italy; LBT Beteiligungsgesellschaft, Germany, representing the Max-Planck Society, The Leibniz Institute for Astrophysics Potsdam, and Heidelberg University; The Ohio State University, and The Research Corporation, on behalf of The University of Notre Dame, University of Minnesota and University of Virginia.
Keck is operated as a scientific partnership between the California Institute of Technology, the University of California, and NASA. Keck access was provided by NASA, the California Institute of Technology, and Northwestern University and the Center for Interdisciplinary Exploration and Research in Astrophysics (CIERA). We are deeply indebted to those of Hawaiian ancestry on whose sacred mountain we are privileged to be guests. We also thank the staff of the W. M. Keck Observatory for their dedicated efforts in maintaining and operating the instruments and telescopes.
\end{acknowledgements}
\begin{facilities}
Keck I (MOSFIRE), JWST (NIRSpec), LBT (MODS)
\end{facilities}

\begin{software}
Cue \citep{2025Li}, Matplotlib \citep{2007Hunter}, NumPy \citep{2020Harris}, pandas \citep{2025Pandas}, PyNeb \citep{2015Luridiana}, SciPy \citep{2025Gommers}
\end{software}

\bibliography{sample701}{}
\bibliographystyle{aasjournalv7}


\end{CJK*}
\end{document}